\documentclass[fleqn,usenatbib]{mnras}

\usepackage{newtxtext,newtxmath}
\usepackage[T1]{fontenc}
\usepackage{ae,aecompl}

\usepackage{graphicx}	% Including figure files
\usepackage{amsmath}	% Advanced maths commands
\usepackage{amssymb}	% Extra maths symbols
\usepackage{tikz}       % Producing diagrams, e.g. Figure 1.
\usepackage{amsmath}    % Math symbols
\usepackage{siunitx}    % Nice formatting of SI units 
\usepackage{multirow}   % Multiple rows in tables
\usepackage{natbib}     % 
\newcommand{\dll}{\Delta \lambda/\lambda}  % delta lambda over lambda
\newcommand{\dff}{\Delta f/f}              % delta frequency over frequency
\newcommand{\mps}[1]{\SI{#1}{\meter\per\second}}    % meters per second
\newcommand{\cmps}[1]{\SI{#1}{\centi\meter\per\second}}     % centimeters per second
\newcommand{\ghz}[1]{\SI{#1}{\giga\hertz}}       % gigahertz
\newcommand{\secref}[1]{section~\ref{#1}}       % refer to section
\newcommand{\figref}[1]{Figure~\ref{#1}}        % refer to figure
\newcommand{\tabref}[1]{Table~\ref{#1}}         % refer to table

\DeclareMathOperator{\erf}{erf} % Adds the error function as a math symbol
\title[Precision and consistency of astrocombs]{Precision and consistency of astrocombs}

\author[D. Milakovi\'c et al.]
{Dinko Milakovi\'c$^{1}$\thanks{E-mail: dmilakov@eso.org},
Luca Pasquini$^{1}$,
John K. Webb$^{2}$,
Gaspare Lo Curto$^{1}$
\\
$^{1}$European Southern Observatory, Karl-Schwarzschild-Str. 2, 85748 Garching bei M\"unchen, Germany\\
$^{2}$University of New South Wales Sydney, Sydney NSW 2052, Australia\\ 
}

\date{Accepted 2020 January 31. Received 2020 January 24; in original form 2019 September 26}

\pubyear{2020}

\begin{document}
\label{firstpage}
\pagerange{\pageref{firstpage}--\pageref{lastpage}}
\maketitle

\begin{abstract}
Astrocombs are ideal spectrograph calibrators whose limiting precision can be derived using a second, independent, astrocomb system. We therefore analyse data from two astrocombs (one \ghz{18} and one \ghz{25}) used simultaneously on the HARPS spectrograph at the European Southern Observatory. The first aim of this paper is to quantify the wavelength repeatability achieved by a particular astrocomb. The second aim is to measure wavelength calibration consistency between independent astrocombs, that is to place limits or measure any possible zero-point offsets. We present three main findings, each with important implications for exoplanet detection, varying fundamental constant and redshift drift measurements. Firstly, wavelength calibration procedures are important: using multiple segmented polynomials within one echelle order results in significantly better wavelength calibration compared to using a single higher-order polynomial. Segmented polynomials should be used in all applications aimed at precise spectral line position measurements. Secondly, we found that changing astrocombs causes significant zero-point offsets ($\approx \cmps{60}$ in our raw data) which were removed. Thirdly, astrocombs achieve a precision of $\lesssim \cmps{4}$ in a single exposure ($\approx 10\% $ above the measured photon-limited precision) and \cmps{1} when time-averaged over a few hours, confirming previous results. Astrocombs therefore provide the technological requirements necessary for detecting Earth-Sun analogues, measuring variations of fundamental constants and the redshift drift.

\end{abstract}

%%%%%%%%%%%%%%%%%%%%%%%%%%%%%%%%%%%%%%%%%%%%%%%%%%

%%%%%%%%%%%%%%%%% BODY OF PAPER %%%%%%%%%%%%%%%%%%

\section{Introduction}

Measuring spectroscopic velocity shifts ($\Delta \lambda/\lambda$) in high resolution astronomical spectra is a powerful and widely used tool in a range of astronomical disciplines. It is used to detect planets outside of the Solar System \citep{Mayor1995,AngladaEscude2016}, look for variation in the values of fundamental constants \citep[e.g. the fine strucure constant, $\alpha$,][ and others]{Dzuba1999,Webb1999,Webb2011,King2012}. It is also the method proposed to measure the expansion of the Universe in ``real time'' and in a model-independent way (also known as the Sandage test or the ``redshift drift'' measurement, \citet{Sandage1962,Loeb1998,Liske2008}) and to map the gravitational potential of the Galaxy \citep{Ravi2019,Leao2019}. These science goals are important drivers for all 30-meter class telescopes planned for the 2020s, i.e. the Extremely Large Telescope \citep[ELT,][]{Tamai2018}, the Thirty Meter Telescope \citep[TMT,][]{Simard2016}, and the Giant Magellan Telescope \citep[GMT,][]{Fanson2018}. This is why all will be equipped with a high resolution optical spectrograph. 

The success of these projects relies not only on increasing the light gathering capability of telescopes, but also on the instrument stability, and the precision and accuracy with which velocity shifts in astronomical spectra can be measured. In this context, precision is the repeatability of subsequent wavelength measurements to each other and accuracy is the closeness of a measured wavelength to its true value. The most demanding of the four aforementioned projects -- the redshift drift measurement -- requires velocity shifts be measured with a precision better than 3 part in 100 billion over spectral ranges of several hundred nanometers and a period longer than ten years. This corresponds to $\dll = 3\times 10^{-11}$, or equivalently, 1 centimeter per second (\cmps{}). The best way to ensure long term precision is to ensure that the measured wavelengths are also accurate. Instruments intended to perform these experiments are therefore designed to have wavelength calibration precision around \cmps{1} and accuracy of order 1 metre per second (\mps{}) over a period of a decade \citep[e.g.][]{Liske2014,Marconi2016}. The success of these projects thus critically relies on the precision and accuracy of the wavelength calibration reaching these levels. 

The currently most commonly used method of wavelength calibration in high resolution spectrographs uses hollow cathode lamp, most commonly Thorium (Th) and Uranium (U) lamps. The hollow cathode lamp calibration suffers from several major drawbacks limiting its precision to a few tens of \cmps{} over a period of one year and average accuracy of approximately \mps{1} in the same period \citep{Lovis2006}, thus falling short of the previously stated goals. It is now generally accepted that laser frequency comb system \citep[LFC,][]{Udem2002,Haensch2006} can achieve substantially better results to push the precision and accuracy of wavelength calibration to the \cmps{1} level and ensure the feasibility of the scientific projects mentioned above. For a recent review about laser frequency comb spectroscopy, see \citet{Haensch2019}.

Laser frequency combs offer significant advantages over arc lamps: (i) they produce thousands of unblended emission lines of uniform intensity and equidistant in frequency; and (ii) the frequencies of LFC lines are {\it a priori} known with accuracy of the atomic clock to which the system is coupled (typically $\Delta f / f \approx 10^{-11}$ or \SI{3}{\milli\meter\per\second}). \citet{Murphy2007b} discusses the advantages of LFCs in more detail. LFCs do not suffer from lamp aging in the same way as hollow cathode lamps, but their components (e.g. the photonic crystal fibre, PCF) degrade. 

The European Southern Observatory (ESO) formed a consortium to develop and install an astronomical LFC (also known as an ``astrocomb'') on the HARPS instrument in 2008. The development saw several test campaigns after which the astrocomb was permanently installed on HARPS in May 2015. This astrocomb has already demonstrated short-term precision at the photon noise level, $\dll=6\times10^{-9}$ or $\cmps{2.5}$ \citep{Wilken2012}. Similar short-term precision using astrocombs has also been demonstrated on several other high-resolution spectrographs \citep[e.g.][]{Ycas2012,Phillips2012,Doerr2012,Glenday:15,Brucalassi2016,McCracken2017}. All these previous studies referenced the astrocomb to itself, so any possible systematic effects arising in the astrocomb system itself may go undetected.

Definitive proof of astrocomb performance can only come from its comparison to another calibration source of the same (or higher) precision. This is why, in a campaign that took place in April 2015, two independent astrocomb systems were installed on HARPS: one constructed for the HARPS instrument itself and the other constructed for the FOCES instrument \citep{Pfeiffer1992}. The experiment had two goals. The first one was to understand whether the precision of a single astrocomb is confirmed by an independent system, and the second one was to assess zero-point offsets in the wavelength calibration introduced by switching between the two astrocombs.

\citet{Probst2020} analyses the same dataset, but with a focus on describing the experimental setup, astrocomb hardware and its optimisation during the campaign before the astrocomb's deployment in May 2015. This paper, on the other hand, focuses on data analysis techniques and advanced algorithms that will provide wavelength calibration precision required by the ELT projects discussed above. 

The paper is divided as follows: in \secref{sec:experiment} we briefly describe the experimental setup -- the spectrograph and the two astrocombs. \secref{sec:data} describes the dataset and general properties of the spectra. \secref{sec:lines} gives details on our automatic algorithm to detect astrocomb lines and obtaining their centres and wavelengths. Wavelength calibration is discussed in \secref{sec:wavecal}, where we describe how we deal with instrumental effects that impact significantly on wavelength calibration precision and accuracy, e.g. defects associated with the CCD manufacturing process. In the same section, we describe our findings on the optimal wavelength calibration model for HARPS spectra. We present our findings on precision and consistency in \secref{sec:analysis}, where we also consider and model the contribution of flux dependent velocity shifts. Finally, our results are presented in \secref{sec:results} and discussed in \secref{sec:discussion}.

\section{Experimental setup}
\label{sec:experiment}

\subsection{The HARPS instrument}
\label{subsec:harps} 
The High Accuracy Radial velocity Planet Searcher \citep[HARPS,][]{Mayor2003} was built for extreme stability and precision and is one of the most stable astronomical spectrographs in existence. HARPS is a fibre fed, high-resolution ($R=\lambda/\Delta\lambda=115000$), R4 grism cross-dispersed echelle spectrograph installed on the 3.6m telescope at ESO's La Silla Observatory and the first instrument to be equipped with an astrocomb for regular operations \citep{Probst2016}. The light of the two input fibres is dispersed into 72 echelle orders on the detector, simultaneously covering the wavelength range between 378 and 691 nm. The spectrograph is enclosed in a thermal and pressure controlled vacuum vessel, with long-term temperature variations at the 0.01K level and operating pressure below $\SI{1e-3}{\milli\bar}$. 

Several thermal and mechanical effects can slightly shift the positions of the spectrum on the detector with time, an unavoidable effect that is eliminated through simultaneous referencing: drifts in the science fibre are tracked by simultaneously observing a spectrum rich in velocity information content (e.g. ThAr, an astrocomb, or a Fabry-P\'{e}rot etalon) in the secondary fibre \citep{Baranne1996}. Each fibre has a static double scrambler. A servo controller (``secondary guiding'') ensures that the object image is always centered in the object fibre. In order to ensure light entrance stability and proper mode mixing in the fibres, a dynamical fibre scrambler that shakes the fibres was added to the setup, adding a temporal scrambling of light \citep{Probst2020}. 

\subsubsection{The detector}
\label{subsubsec:detector}
The HARPS detector is a mosaic of two EEV2k4 CCDs (red and blue). Each CCD is constructed by stitching together $1024\times512$\,pix$^2$ segments: eight in the dispersion ($x$) direction and two in the cross-dispersion ($y$) direction. CCD pixels have a nominal size of $15\times15 \mu\rm{ m^2}$ but errors in the segment stitching process results in deviations from the nominal pixel size at segment boundaries, i.e. every 512 pixels in the $x$ direction and every 1024 in the $y$ direction. 

The HARPS detector was the first one shown to suffer from the effect of imperfect CCD stitching \citep{Wilken2010}. \citet{Dumusque2015} showed that lines which cross segment boundaries produce spurious velocity shifts as high as a few \mps{} with a period of one year in HARPS observations. Proper mitigation of this effect is therefore important for exoplanet detection and also for fundamental constant and redshift drift measurements. The way this has been done previously is to shift the measured positions of calibration lines (in pixel space) by the measured size of pixel size anomalies during wavelength calibration. Pixel size anomalies were measured by \citet{Bauer2015} using Fabry-P\'{e}rot etalon exposures and by \citet{Coffinet2019} using flat-field exposures. We perform a similar measurement using astrocombs exposures in \secref{subsec:gaps}. Therefore, in \secref{subsec:wavesol}, we examine the effectiveness of different calibration methods in removing the effect of pixel size anomalies.

\subsection{The astrocombs}
\label{subsec:astrocombs}
Astrocombs are laser frequency comb systems built specifically to wavelength calibrate astronomical spectrographs \citep{Steinmetz2008}. They produce thousands of emission lines (or modes) of uniform intensity with precisely known frequencies. The nominal frequency of each mode is given by the ``LFC equation'':
\begin{equation}
    f_n = f_{o} + n\times f_{r},
    \label{eq:lfc}
\end{equation}
where $f_{o}$ and $f_{r}$ are the ``offset'' and the ``repetition'' frequencies, and $n$ is the mode number (a large positive integer). Both $f_o$ and $f_r$ are radio-frequencies referenced to an atomic clock and known with precision of $\dff = 5.6\times 10^{-12}$ over the timescale of several hours \citep{Probst2020}. The frequency, and the wavelength, of each line can therefore be determined with the same precision. 

The HARPS astrocomb development saw several test campaigns between January 2009 and April 2015. One of the goals of the April 2015 campaign was to characterise the performance of the HARPS astrocomb against a completely independent second one. The second astrocomb, built for the FOCES instrument \citep{Pfeiffer1992}, was loaned from the Wendelstein Observatory (operated by Ludwig-Maximillians-Universit\"at, LMU) for this purpose. The HARPS astrocomb has an \ghz{18} line separation. Since the HARPS spectral resolution is around \ghz{5} in the middle of its spectral range, astrocomb lines are kept well apart with virtually no residual overlap. We will refer to this \ghz{18} astrocomb as ``LFC1'' in further text. To accommodate the lower resolution of the FOCES instrument ($R=70 000$), the FOCES astrocomb was designed with a wider line separation of \ghz{25}. We will refer to this astrocomb as ``LFC2'' in further text. Wavelength coverage of the two astrocombs differs slightly due to different requirements for the HARPS and FOCES instruments. Relevant information about the two astrocombs are tabulated in \tabref{tab:setup}. See \citet{Probst2020} for a more comprehensive description of the astrocomb design and the setup during the April 2015 campaign.

\section{Data}
\label{sec:data}
\subsection{The dataset}
\label{subsec:dataset}
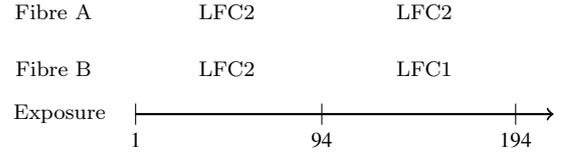
\begin{figure}
    \centering
    
    \begin{tikzpicture}

        \node (A) at (-0.08,1.35)  {Fibre A};
        \node (B) at (-0.08,0.6)   {Fibre B};
        
        \node (A1) at (2.2,1.35) {LFC2};
        \node (B1) at (2.2,0.6)  {LFC2};
        \node (A2) at (4.8,1.35) {LFC2};
        \node (B2) at (4.8,0.6)  {LFC1};
        \node (L)  at (0,0)      {Exposure};
        \coordinate [] (start) at (1,0);
        \coordinate [] (end) at (6.5,0);
        \draw[thick,->] (start) -- (end) node[anchor= north west] {};
        %\foreach \x in {0,2.9,6}
        \draw (1.0,4pt) -- (1.0,-4pt) node[anchor=north] {$1$};
        \draw (3.45,4pt) -- (3.45,-4pt) node[anchor=north] {$94$};
        \draw (6.0,4pt) -- (6.0,-4pt) node[anchor=north] {$194$};

    \end{tikzpicture}
    \caption{We analyze a time series of 194 astrocomb exposures. Fibre A carrying LFC2 light was used for simultaneous referencing throughout the series. Fibre B carried LFC2 light during the first 94 exposures, after which it was switched to carry LFC1 light. We measure velocity shifts of all exposures, compensating for unavoidable spectrograph drifts, in order to establish the precision of each astrocomb and the consistency of their wavelength calibrations.}
    \label{fig:setup}
\end{figure}

We use a time series of spectra of the two astrocombs described above for our analysis. The series consists of a total of 194 exposures. Each exposure was 30 seconds long with a read-out time of 22 seconds. The entire duration of the series is approximately six hours, with a two hour gap between the end of exposure 94 and the beginning of exposure 95. In the first 94 exposures, fibre B (the object fibre) was illuminated with LFC2, after which the astrocomb was changed and 100 exposures of LFC1 were taken in the same fibre. We therefore divide the dataset into two samples, depending on which astrocomb illuminated fibre B. We will refer to the ``LFC2 sample'' for exposures 1--94 and to the ``LFC1 sample'' for exposures 95--194. Meanwhile, LFC2 was used as a simultaneous reference in fibre A (simultaneous fibre), keeping track of spectrograph drifts throughout the whole series, for a total of 194 exposures in fibre A. \figref{fig:setup} gives a schematic of the series of exposures we use in our analysis. 

\begin{table}
    \centering
    \begin{tabular}{ccc}
         & LFC1 & LFC2 \\
    \hline
    $f_o$           & \ghz{5.7}                 & \ghz{9.27}  \\
    Native $f_r$           & \SI{250}{\mega\hertz}     & \SI{250}{\mega\hertz} \\
    Mode filtering  & 72                        & 100 \\
    $f_r$ & \ghz{18}                  & \ghz{25} \\
    $\lambda_{min}$ & \SI{438.8}{\nano\meter}   & \SI{455.4}{\nano\meter} \\
    $\lambda_{max}$ & \SI{691.5}{\nano\meter}   & \SI{691.5}{\nano\meter} \\
    \end{tabular}
    \caption{Basic parameters of astrocombs LFC1 and LFC2. The two share the basic design but have been optimised for different instruments. See \citet{Probst2020} for a more comprehensive description.}
    \label{tab:setup}
\end{table}

\begin{figure}

	\includegraphics[width=\columnwidth]{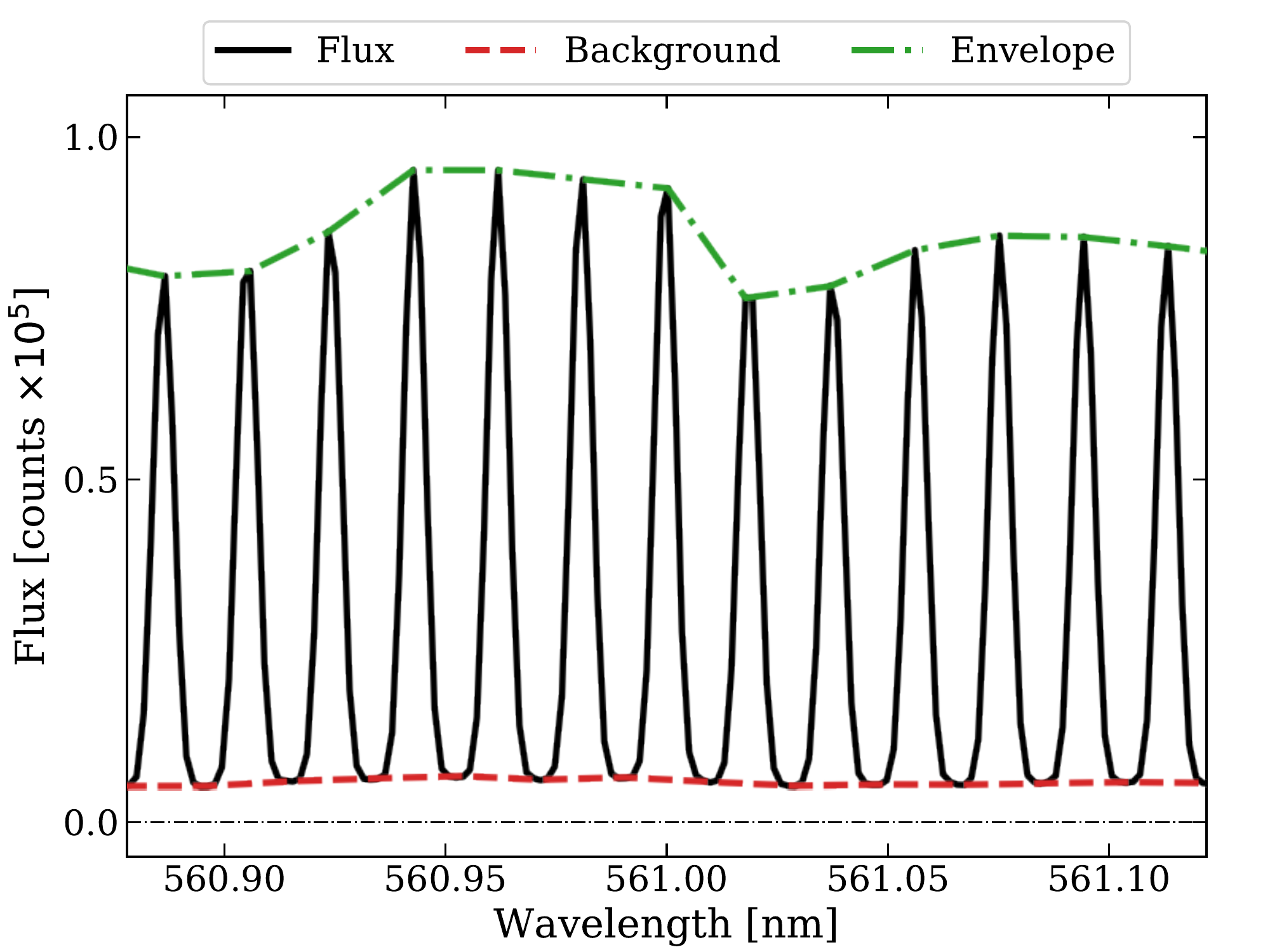}
   \caption{Zoom-in of the region between 560.8 and 561.1 nm of the 1D extracted spectrum of LFC1 in echelle order 109 (solid black line). Signal amplification gives rise to the background (dashed red line), which contributes approximately 13\% of the total flux in a single exposure, as calculated from its ratio to the envelope (dot-dashed green line). The background is removed before line fitting. }
    \label{fig:spec_lfc1}
\end{figure}

We choose to work on `{\sc e2ds}' files -- unmerged 1D spectra extracted from raw images by the HARPS pipeline (version v3.8) using optimal extraction after Horne (1986). Each {\sc e2ds} file consists of 72 (71) echelle orders covering 4096 pixels of an exposure in fibre A (B). The detector covers echelle orders 89 to 161, with the exception of order 115 for fibre A and orders 115 and 116 for fibre B, which fall in between the two detector CCDs. A fraction of the LFC1 spectrum showing individual lines around $\SI{561}{\nano\meter}$ in echelle order 109 is plotted in \figref{fig:spec_lfc1}.

\begin{figure}
	\includegraphics[width=\columnwidth]{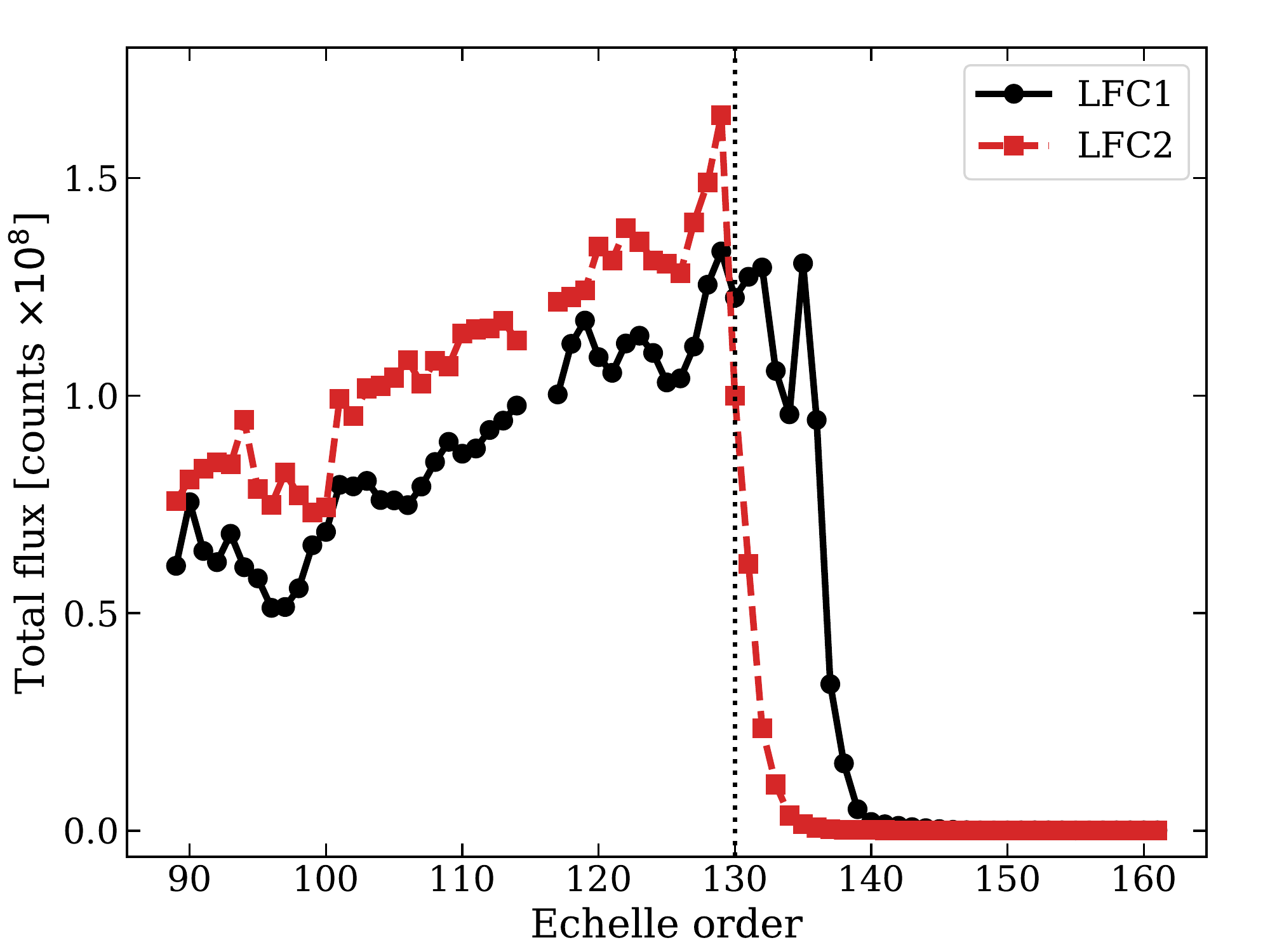}
   \caption{The light intensity of the two astrocombs in fibre B drops sharply above order 136 for LFC1 (black circles) and above order 130 for LFC2 (red squares). The two astrocombs cover more than 70\% of the total HARPS wavelength range. We limit our analysis to orders 89--130, where the two astrocombs have sufficient and comparable flux. Orders 115 and 116 fall in between the two detector CCDs and therefore have no measured flux.}
    \label{fig:cts_per_ord}
\end{figure}

We limit our analysis to echelle orders 89--130, where the fluxes of the two astrocombs are sufficiently high and comparable -- as evidenced by the total number of counts detected in each echelle order (see \figref{fig:cts_per_ord}). This covers wavelengths between $\SI{468.1}{\nano\meter}$ and $\SI{691.5}{\nano\meter}$, or 70\% of the total HARPS wavelength range. 

A quantity that is directly relevant for exoplanet detection studies, but not for varying constant or redshift drift measurements, is the uncertainty on the mean velocity shift that can be determined in a spectrum. This measure of uncertainty is determined by the photon noise and other detailed spectral attributes -- see \citet{Bouchy2001} and \citet{Murphy2007b} for details. We will refer to this quantity as ``photon-limited velocity precision''. We calculate the photon-limited velocity precision of individual exposures in our dataset across orders 89 -- 130. We find that the average photon-limited velocity precision of a single exposure in the LFC1 sample is \cmps{2.3} (\cmps{3.0}) in fibre A (B); and in the LFC2 sample is \cmps{2.3} (\cmps{2.5}) in fibre A (B). The reduced photon-limited velocity precision in fibre B is attributable to lower flux compared to fibre A for both samples.

\subsection{Spectral background}

The spectra obtained using both astrocombs exhibit a strong background component (the red line in \figref{fig:spec_lfc1} illustrates the background light in the LFC1 spectrum). This background originates mostly from amplified spontaneous emission (ASE) in the high-power Yb-fibre amplifier \citep{Probst2013,Probst2020} and, to a lesser extent, from scattered light \citep[measured to be $<1\%$ at \SI{590}{\nano\meter} in the HARPS spectrograph][]{HARPS-manual}. The ASE from the high-power amplifier is further amplified by non-linear processes in the photonic crystal fibre, located just after it in the optical path. %Exchanging the photonic crystal fibre, leaving all other astrocomb system components unchanged, caused the background to change by a factor of a few. 
We refer the reader to \citet{Probst2020} for more technical details on the origin of the background.

The background is highly modulated, approximately tracing the variations in intensity of astrocomb lines. We characterize the contribution of the background to the total detected flux in terms of the background-to-envelope ratio (B2E). The background here refers to a function connecting the local minima in the spectrum. This is a piece-wise linear function of pixel number, $x$:
    \begin{equation}
        B(x)\Bigr|_{x_1}^{x_2} = F(x_1) + \frac{F(x_2)-F(x_1)}{x_2 - x_1} \times (x_2 - x)
    \end{equation}
where $F$ is flux (in units counts) and $x_1$ and $x_2$ are locations of adjecent minima in the spectrum. The envelope is the analogous quantity except maxima instead of minima are fitted. Both are illustrated in Figure \ref{fig:spec_lfc1}.

Our analysis shows that B2E increases linearly with decreasing wavelength, with values between 2\% and 16\% in a typical exposure in both astrocombs. The background is therefore non-negligible. To minimise any impact on estimating astrocomb line centres, we subtract the background from the total flux, propagating the errors to correctly modify the spectral variance array:
\begin{equation}
    \sigma^2(x) = F(x) + B(x),
\end{equation}
where we have assumed that the flux and the background are Poissonian and the detector noise (i.e. dark-current, read-out-noise) is negligible. 

% -------------------------------------------------------------------------------------------
\section{Astrocomb lines}
\label{sec:lines}

\subsection{Line detection}
Our line detection algorithm automatically detects all astrocomb lines in the exposure. The algorithm relies on locating minima between individual lines as their natural limits. We found this practice to be preferred over using maxima, as the latter caused issues with falsely detected and skipped lines when using an automated detection routine. Line detection is done in three steps. In the first step, we smooth the recorded spectrum of an echelle order with a Wiener filter with a 3 (5) pixel wide window for LFC1 (LFC2). The smoothed spectrum makes identifying minima easier in the following step. In the second step, we identify local minima as points where the first derivative (with respect to the pixel number) switches sign, and the second derivative is larger than zero. This step sometimes falsely detects minima in the data, especially when the signal-to-noise ratio (S/N) is low. We therefore reject falsely detected minima in the third and final step, using the following two assumptions: (i) the distance between minima must not significantly deviate from typical distance between astrocomb lines; and (ii) the distance between minima within the echelle order increases approximately linearly with increasing wavelength.

We use the first assumption to remove minima closer together than 90\% of the typical distance between lines in the same echelle order, where the latter is equal to period of the strongest peak in the periodogram of the order. Depending on the astrocomb and the echelle order, this number is between 11 and 20 pixels, with LFC2 always having larger values due to larger mode separation. The second assumption follows directly from Equation~ \eqref{eq:lfc}: the separation of consecutive lines in wavelength space approximately follows $\sim\lambda_n/n$, where $\lambda_n$ is the wavelength of the $n^{\rm th}$ astrocomb mode. This means that the distance between lines increases approximately linearly in pixel space within a single echelle order. We therefore remove $>3\sigma$ outliers to the linear function best describing $\Delta x (x)$, where $\Delta x$ is the distance between adjacent lines. 

Our automatic line detection algorithm detects $\approx 13300$ and $\approx 9800$ in each LFC1 and LFC2 exposure (where the background was subtracted), respectively. Across the entire dataset, we detect $N_A=1898254$ lines in fibre A and $N_B=2222168$ lines in fibre B, with a total $N = 4120422$ lines. The average signal-to-noise ratio (S/N) of lines is approximately 165, with $\approx 10\%$ differences between LFC1 versus LFC2 and fibre A versus fibre B.

% -------------------------------------------------------------------------------------------
\subsection{Profile fitting}
\label{subsec:gauss}

In this analysis we assume that each astrocomb line can be well-represented by a single Gaussian profile. Visual examination of the data suggests that the approximation is generally reasonable although asymmetries are seen depending on position on the CCD. We will explore different ways of modelling astrocomb profiles in a forthcoming paper.

\begin{figure}
    \centering
    \includegraphics[width=\linewidth]{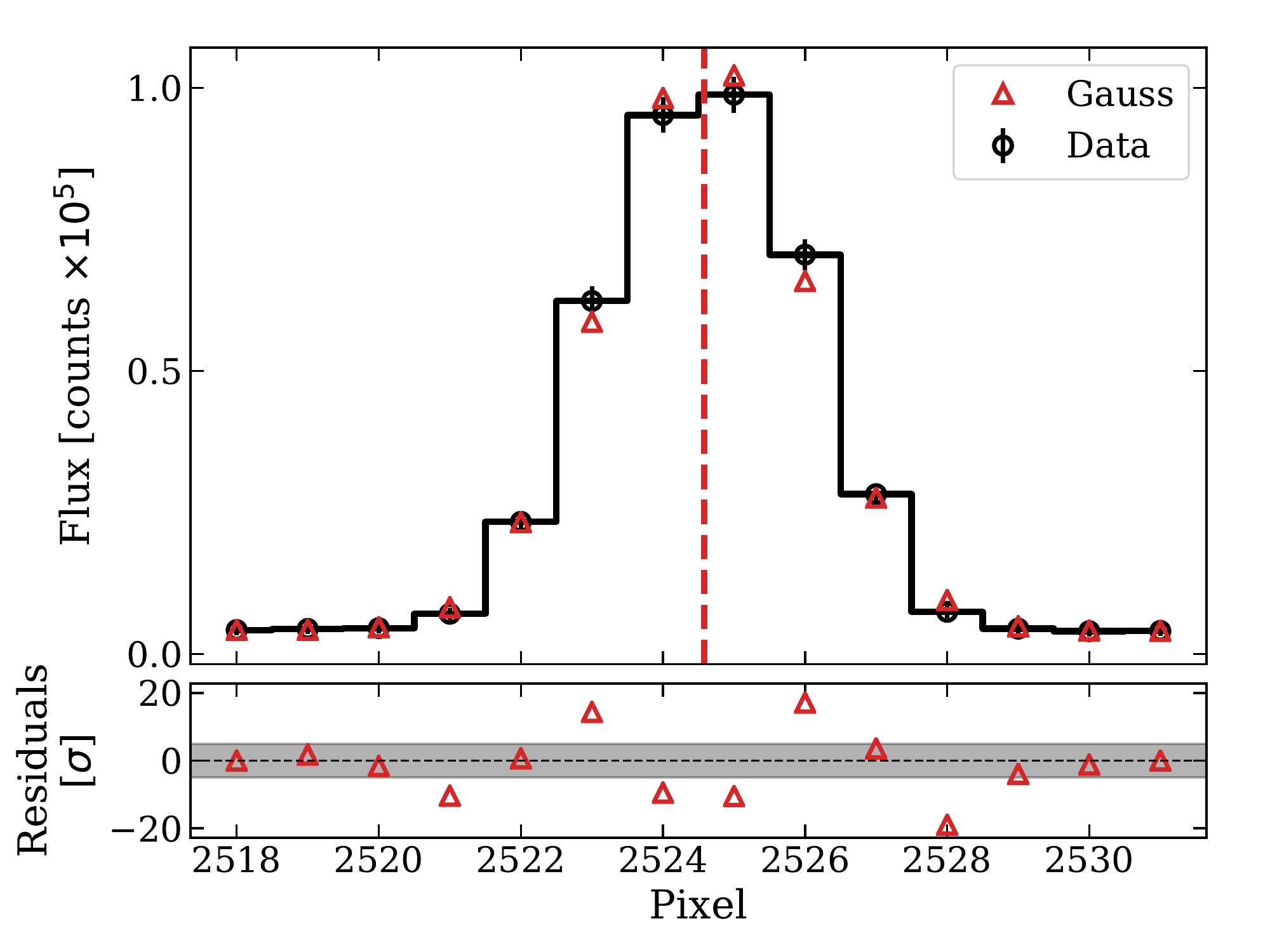}
    \caption{We fit Gaussian profiles to astrocomb lines after spectral background removal. The Gaussian function is integrated under each pixel during fitting. A smooth Gaussian curve is therefore not plotted. {\it Top}: The data (black circles) and the Gaussian model (red triangles). The best line centre estimate is given by the vertical red dashed line. Errors on the data are enlarged by a factor of 10 for visibility. {\it Bottom}: Normalised residuals show the Gaussian model is ultimately not the correct line profile. The shaded gray area shows the $5\sigma$ range.}
    \label{fig:linefit}
\end{figure}

We use a Levenberg-Marquardt algorithm\footnote{\texttt{leastsq} routine from \texttt{scipy.optimize}} to derive best-fit Gaussian parameters for each line. Since the data are high signal-to-noise, and since slight non-linearities in the model function exist across individual spectral pixels, the model-fitting procedure avoids simply computing the Gaussian value at the centre of each pixel but instead performs an integration of the flux falling within each pixel. The expected number of counts in each pixel (with $x_L$ and $x_R$ its left and right boundaries) is given by:
    \begin{equation}
        \Phi(x_L,x_R) = A \, \sigma \, \sqrt{\dfrac{\pi}{2}} \left[\erf\left(\frac{x_R-\mu}{\sqrt{2}\sigma}\right) -\erf\left(\frac{x_L-\mu}{\sqrt{2}\sigma}\right)\right].
    \end{equation}
Here $\erf$ is the error function, and $A$, $\mu$, and $\sigma$ are the amplitude, the mean and the standard deviation of the Gaussian, respectively. The algorithm provides a line-centre uncertainty estimate for each line. The mean line-centre uncertainty across all $\approx4{\rm M}$ detected lines is 3 milli-pixel (mpix).
An example fit for a single LFC1 line is shown in \figref{fig:linefit}.

\begin{figure*}
	\includegraphics[width=0.8\textwidth]{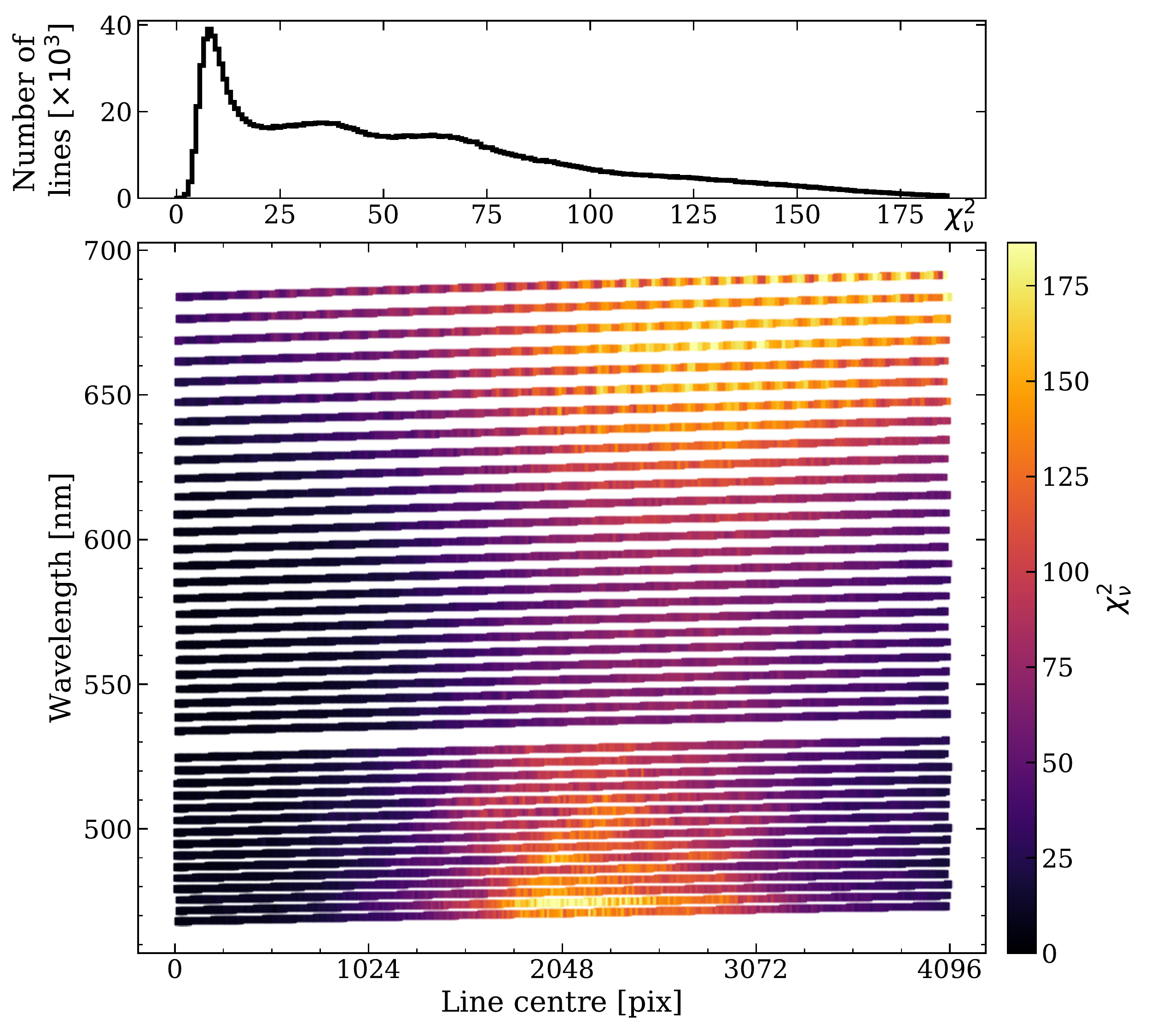}
   \caption{{\it Main panel}: The distribution of $\chi_\nu^2$ values for the Gaussian profile fits shows that a Gaussian profile does not provide a satisfactory fit to the data. There exists a concentration of high $\chi^2_\nu$ values in two places on the detector, showing that the astrocomb line profile changes systematically across the detector. {\it Top panel}: The histogram of values in the main panel, total number of lines $N = 1898254$. The distribution peaks at $\chi_\nu^2=7.9$.}
    \label{fig:CCD_gauss_mean_chisq}
\end{figure*}

Ultimately, the Gaussian approximation above is incorrect. This can be seen by eye from the asymmetric shape of astrocomb lines and is evident from the high values of reduced $\chi^2$ values ($\chi^2_\nu = {\chi^2}/{\nu}$, with $\nu$ the number of degrees of freedom) we get for the Gaussian line fits. The $\chi^2_\nu$ distribution derived from Gaussian fitting, shown in \figref{fig:CCD_gauss_mean_chisq}, peaks at $\chi_\nu^2 = 7.9$ and has a mean $57.6$ across the detector, indicating an overall poor fit to the data. Large $\chi^2_\nu$ values for Gaussian fits are concentrated in two regions: in the red half ($x>2048$) of echelle orders 89--98 and the middle part of orders 125--130 ($x\approx2048$). The $\chi^2_\nu$ values do not correlate with any of the fit parameters or their errors. The same pattern is seen independently for both astrocombs and in both fibres. This suggests that the $\chi^2_\nu$ pattern must be due to variation in the line-spread function profile across the detector. 

Whilst a Gaussian is clearly not the correct line shape, we show later that the Gaussian approximation nevertheless performs well in term of repeatability such that radial velocity studies are relatively unaffected by the correlated patterns seen in \figref{fig:CCD_gauss_mean_chisq}. We note however that this issue will be important for other types of studies such as varying constants and redshift drift. We will later derive a model of the line-spread function -- assuming it will also give us a more accurate estimation of the line centre \citep{MilakovicPrep}.

\subsection{Mode identification}
Astrocomb wavelength calibration requires another wavelength calibration source to establish the absolute scale by identifying a single line in each echelle order, after which the frequencies of all lines are known by counting. This is not a critical aspect as long as the error in the absolute calibration source is much smaller than half of the separation between astrocomb lines. The local accuracy of ThAr wavelength calibration, the HARPS standard, is between $10$ and \mps{80} or $\Delta f = 0.02$ to \ghz{0.15} at \SI{550}{\nano\meter}, (respectively). This is well below the line separation of either astrocomb (\tabref{tab:setup}). We therefore use the ThAr wavelength calibration to identify a single astrocomb mode per echelle order in the following way. The mode number, $n$, is the nearest integer to the number:
    \begin{equation}
        \label{eq:mode}
        n = {\rm nint} \Big (\frac{f_n^{ThAr} - f_o}{f_r} \Big ),
    \end{equation}
where $f_n^{ThAr}$ is frequency of the astrocomb line determined from the ThAr wavelength calibration, $f_o$ and $f_r$ are the astrocomb offset and repetition frequencies, respectively. These frequencies are recorded by the astrocomb system\footnote{The experimental nature of the April 2015 campaign meant $f_o$ of both astrocombs was changed multiple times. We discover an un-noted shift in LFC1 of \SI{100}{\mega\hertz} whilst analysing the daset. More details can be found in Appendix \ref{app:harps_shift}.}. We always use the same ThAr coefficients to determine the mode number of the line closest to pixel 2048, where the ThAr calibration is expected to be the most accurate. 

The wavelength and the corresponding uncertainty of each line are calculated from equations:
    \begin{equation}
        \lambda_{n} = \frac{c}{f_n},
    \end{equation}
    and
    \begin{equation}
        \sigma_\lambda = \frac{c}{f_n^2}\; \sigma_f,
    \end{equation}
with $c$ the speed of light and $\sigma_f$ is the frequency uncertainty for each line. Empirically, $\sigma_f/f = \sigma_{\lambda}/\lambda \approx 10^{-11}$. Uncertainties at this level are generally orders of magnitude below spectral line uncertainties in astronomical targets such as quasars.

% -------------------------------------------------------------------------------------------
\section{Wavelength calibration}
\label{sec:wavecal}
Spectrograph wavelength calibration relates the measured positions of a set of calibration lines on the detector with their known laboratory wavelengths in a way that assigns a wavelength to each position on the detector. The most common approach in optical echelle spectroscopy is to fit a polynomial to a set of calibration lines in each extracted echelle order. The large number of astrocomb lines and the exquisite accuracy with which their wavelengths are known allow us to look for a more realistic model, e.g. by increasing the polynomial order  \citep[see e.g.][]{Wilken2010}. In this Section we examine a range of polynomial orders in an attempt to identify an optimal number of degrees of freedom.

Furthermore, it is generally assumed that all pixels have the same physical size, so that the physical distance between calibration lines on the detector can be expressed in pixel distance. This assumption has been proven invalid by the discovery of the HARPS detector pixel size anomalies (see \secref{subsubsec:detector}). Distortions of the HARPS wavelength scale caused by the pixel size anomalies can be removed using one of the following two approaches: (i) {\it global polynomial}: producing a wavelength calibration spanning the entire echelle order in which the anomalies have been accounted for (see \secref{subsec:gaps}), and (ii) {\it segmented polynomials}: producing a separate wavelength calibration for each 512-pixel CCD segment that an echelle order crosses. \citet{Coffinet2019} take the former approach, whereas \citet{Wilken2010} and \citet{Molaro2013} take the latter. 
In this analysis we use the Akaike information criterion \citep[AIC,][]{Akaike1974}, corrected for the finite sample size \citep[AICc,][]{Sugiura1978}, in order to asses which of these two approaches provides the best results (\secref{subsec:wavesol}).

In what follows, we make extensive use of Weighted Orthogonal Distance Regression\footnote{Python package \texttt {scipy.odr}, based on \texttt{ODRPACK} \citep{ODRPACK}}, an algorithm which allows us to account for both the positional and wavelength uncertainty of astrocomb lines in polynomial fitting.

\subsection{Measuring HARPS detector pixel size anomalies}
\label{subsec:gaps}
We derive our pixel size anomaly measurements from distortions of the HARPS wavelength scale revealed by the astrocomb lines. Astrocomb wavelength coverage limits us to only three out of four $y$-blocks: two blocks of the red CCD (blocks 1 and 2 in further text) and a single block of the blue CCD (block 3 in further text). 

\begin{figure}
    \centering
    \includegraphics[width=\columnwidth]{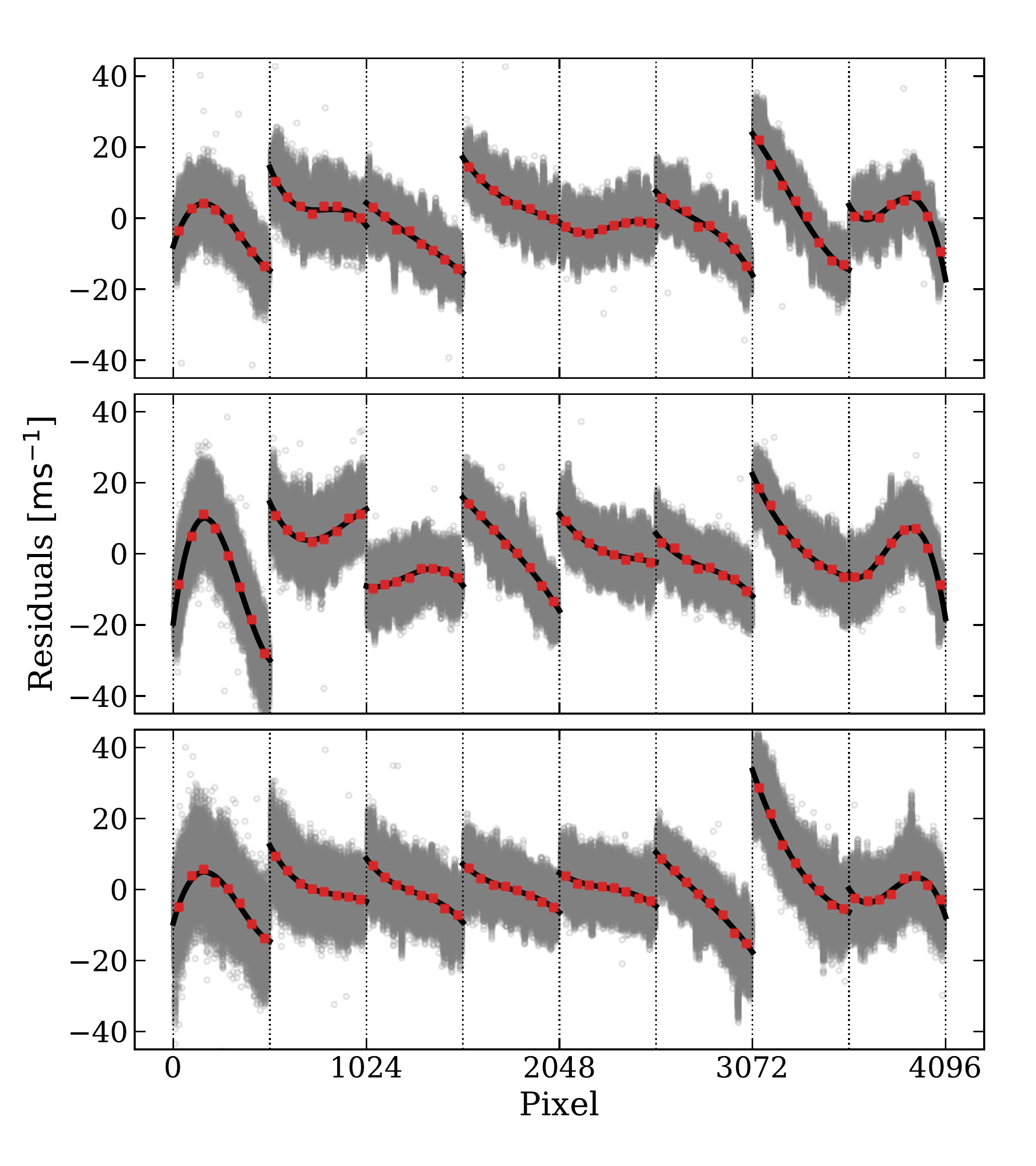}
    \caption{We use $\approx 2{\rm M}$ astrocomb lines detected in fibre A to measure pixel size anomalies imparted during the CCD manufacturing process. The residuals to the global $8^{\rm th}$ order polynomial wavelength solution (gray points) reveal $\approx\mps{20}$ discontinuities the edges of each 512 pixel-wide segment. We bin the residuals into 64 bins and calculate the mean in each (red squares). We fit the means in each segment with a $3^{\rm rd}$ order polynomial and use them to determine the size of the discontinuity. This is measured as the difference between the models (black lines) evaluated at segment borders (dotted vertical lines). The top two panels correspond to blocks 1 and 2 (comprising the red CCD) and the bottom panel corresponds to block 3 (comprising the blue CCD). Block 4 is not illuminated. The average error on each point is $\approx \cmps{5}$ and therefore not visible.
    }
    \label{fig:gaps}
\end{figure}

Pixel size anomalies are measured for each block individually using the wavelength calibration residuals in the following way: 
\begin{enumerate}
    \item For a given detector $y$-block, we consider only those echelle orders which fall onto the block. It seems likely that distortions in the $x$ direction are common to all orders that fall within the same 1024-pixel high CCD block in the $y$ direction. If this is not the case, the effect would be to increase the scatter in the $y$-direction within each $512\times1024$ block (the shaded region) in \figref{fig:gaps};
    \item We fit a global eighth order polynomial to pixel-wavelength pairs of astrocomb lines separately in each echelle order and each exposure. We calculate the residuals to the true line wavelengths (i.e. Equation~ \eqref{eq:lfc}) and express them in \mps{};
    \item We bin the residuals in 64 bins along the $x$ axis (i.e. giving 8 points per 512 pixel-wide segment), excluding lines closer to segment borders than 10 pixels and those with residuals larger than \mps{200};
    \item We fit a third order polynomial to the binned residuals. A typical error on each binned residual is of order \cmps{5} due to the large number of points in each bin ($\approx10-20{\rm k}$).
    \item The discontinuity $g$ is given by the difference between the polynomials in two adjacent segments at their boundary (in units \mps{}):
    \begin{equation}
        g(k) = P_1(x)\Bigr|_{x=k} - P_2(x)\Bigr|_{x=k}.
    \end{equation}
    Here, $P_1$ and $P_2$ are the polynomials in two adjacent segments and $k$ is the position of the discontinuity (in pixels, multiple of 512);
    \item We convert the discontinuity into the pixel size anomaly by dividing it by the size of the HARPS pixel in velocity units in the middle of the HARPS wavelength range: ${\rm 1 pix} = \mps{829}$. 
\end{enumerate}

This is illustrated in \figref{fig:gaps} using data from fibre A. Measurements from fibre B (not illustrated) were found to be consistent with those from fibre A; there are a total of 21 pixel size anomaly measurements from each fibre. Differences between corresponding pixel size anomaly pairs were measured for all 21 pairs. The unweighted mean of those numbers is $\mu=-0.4 \pm 0.4$\,mpix, corresponding to $-0.3\pm\mps{0.3}$. 

Although the two fibres produced completely consistent results, the results from fibre A seemed better than from fibre B, in that the scatter in the fibre B residuals (\figref{fig:gaps}) were more pronounced.  We did not attempt to explore the reason for this and simply used fibre A to make the corrections. These are tabulated in \tabref{tab:gaps} and used to adjust the positions of individual astrocomb lines. The corrected position of one line within one echelle order, in pixels, is given by:
    \begin{equation}
    \label{eq:gaps}
        x_c = x + \sum\limits_k^{k<x} g_k,
    \end{equation}
where $x_c$ and $x$ are the corrected and the fitted line positions of the same line and $g_k$ is the size of a pixel size anomaly located at the $k^{th}$ pixel.

Our measurements are in very good agreement with previous, independent, results. The average agreement with the results of \citet{Bauer2015} is $\mu=3.1\pm1.7$\;mpix ($2.6\pm\mps{1.6}$). Similarly, agreement with the results of \citet{Coffinet2019} is $\mu=2.6\pm0.5$\;mpix ($2.1\pm \mps{0.4}$). This corresponds an agreement between the two measurements at the \SI{40}{\nano\meter} level on the detector. Unlike the flat-field method of \citet{Coffinet2019}, we are not sensitive to the sizes of individual pixels at segment borders but only to the sum of their sizes. Our astrocomb method is, however, complementary to theirs and serves as a consistency check. Furthermore, it demonstrates the usefulness of astrocombs for detector characterization necessary to obtain robust scientific results.

\begin{table}
	\centering
	\caption{The HARPS detector pixel size anomalies are calculated from astrocomb line wavelength residuals obtained using an eighth order global polynomial (see \secref{subsec:gaps} for details). Units are mili-pixels ($1\;\rm{pixel} = 15\mu\rm{m}$).}
	\label{tab:gaps}
	\begin{tabular}{l|cc|cc} 
		
		CCD   & \multicolumn{2}{c}{Red} & \multicolumn{2}{c}{Blue}\\
		\cline{2-5}
		\vspace{0.1cm}
		Orders & 89--99 & 100--114 & 116--134 & 135--161 \\
		Block & 1 & 2 & 3 & 4\\
		\vspace{0.1cm}
		$N_{lines}$  & 423k   & 698k    & 754k   & 0 \\ 
		
		Pixel& \multicolumn{4}{c}{Pixel size anomaly [mpix]}\\
		\hline
		512  & 34.97 & 53.62  & 32.27 & -- \\
		1024 & 7.67  & -23.79 & 13.79 & -- \\
		1536 & 39.12 & 30.88  & 19.72 & -- \\
		2048 & 0.14  & 30.56  & 13.06 & -- \\
		2560 & 11.77 & 9.82   & 18.40 & -- \\
		3072 & 48.28 & 41.86  & 61.89 & -- \\
		3584 & 21.78 & 3.18   & 7.07  & -- \\
		\hline
	\end{tabular}
\end{table}

\subsection{Choosing a wavelength calibration model}
\label{subsec:wavesol}
We return to our aim of determining the model the optimal residuals number of degrees of freedom. We do this using the AIC corrected for small sample sizes \citep[AICc,][]{Sugiura1978} and choose the model providing the smallest residuals using the smallest number of free parameters possible. The AICc is calculated as:
    \begin{equation}
    \label{eq:aicc}
    	{\rm AICc} = \chi^2 + 2\,p + \frac{2p\,(p+1)}{n-p-1},
    \end{equation}
where $p$ is the number of free parameters and $n$ is the number of data points used in the fit. Theoretically, the model with the lowest AICc value is preferred.

We consider a total of 29 wavelength calibration models, grouped into two groups mentioned beforehand: the segmented and the global polynomial models. The segmented polynomial models range between $2^{\rm nd}$ and $12^{\rm th}$ order, whereas the global polynomials range between $3^{\rm nd}$ and $20^{\rm th}$ order. The former have $p=8\times(m+1)$ free parameters and the latter have $p=m+1$ free parameters, with $m$ the highest order polynomial in the model.

For segmented polynomial models, we divide the echelle order into eight 512-pixel wide segments and fit a polynomial in each segment individually. We do not impose conditions on the continuity or smoothness of the polynomials at segment boundaries and leave the parameters in each segment independent, resulting in a discontinuous wavelength calibration model. In the case of global polynomial models, we first adjust the positions of individual lines to account for the CCD stitching pattern using Equation~ \eqref{eq:gaps}, after which we fit a single polynomial to all astrocomb lines in the echelle order. 

\begin{figure}
    \centering
    \includegraphics[width=\columnwidth]{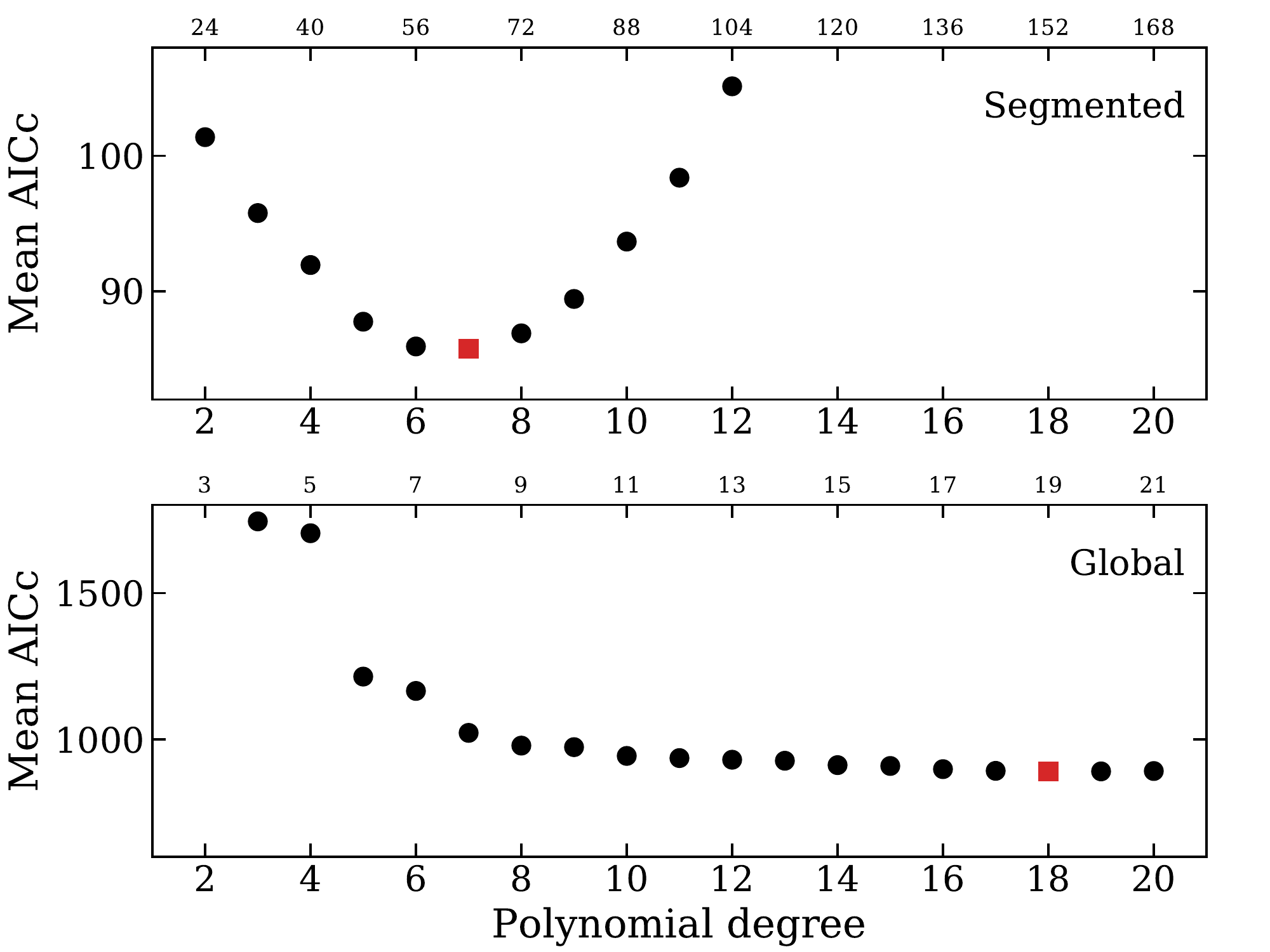}
    \caption{To find the optimal wavelength calibration model, we compute the mean AICc for all 29 models (black circles) divided into two groups: segmented (top panel) and global (bottom panel) polynomials. The preferred model (red square) has the lowest mean AICc value. These are the $7^{\rm th}$ order segmented and the $18^{\rm th}$ order global polynomial models. The top x-axis indicates the total number of free parameters for each model.}
    \label{fig:median_aicc_polynomial}
\end{figure}

We calculate AICc values for all 29 models, for each of the 43 echelle orders, for each of the 194 exposures, and for each fibre. There are therefore 66736 AICc values for each segmented polynomial model (i.e. for each parabolic, cubic, quadratic, quintic, etc.), and 8342 values for each global polynomial model, for each fibre. 
Considering the large number of individual fits, we look at AICc in a statistical sense when comparing models: the best model is the one with the lowest mean AICc, where that mean is averaged over all echelle orders and all exposures. The mean AICc for all 29 models is plotted in \figref{fig:median_aicc_polynomial}, separately for the segmented (top panel) and global polynomials (bottom panel).

The segmented polynomial model with the lowest mean AICc is a seventh order polynomial ($p=56$), whereas the best global polynomial model is an $18^{\rm th}$ order polynomial ($p=19$). 
All global polynomials have AICc values 80-100 times higher than segmented polynomials, indicating that segmented polynomials are preferred in all cases. The AICc retains no spatial information. We therefore look at the histogram of the residuals and explore any possible correlations in the residuals with pixel number for the two wavelength calibration models preferred by AICc: the $7^{\rm th}$ order segmented polynomial and the $18^{\rm th}$ order global polynomial.

\begin{figure}
    \centering
    \includegraphics[width=\columnwidth]{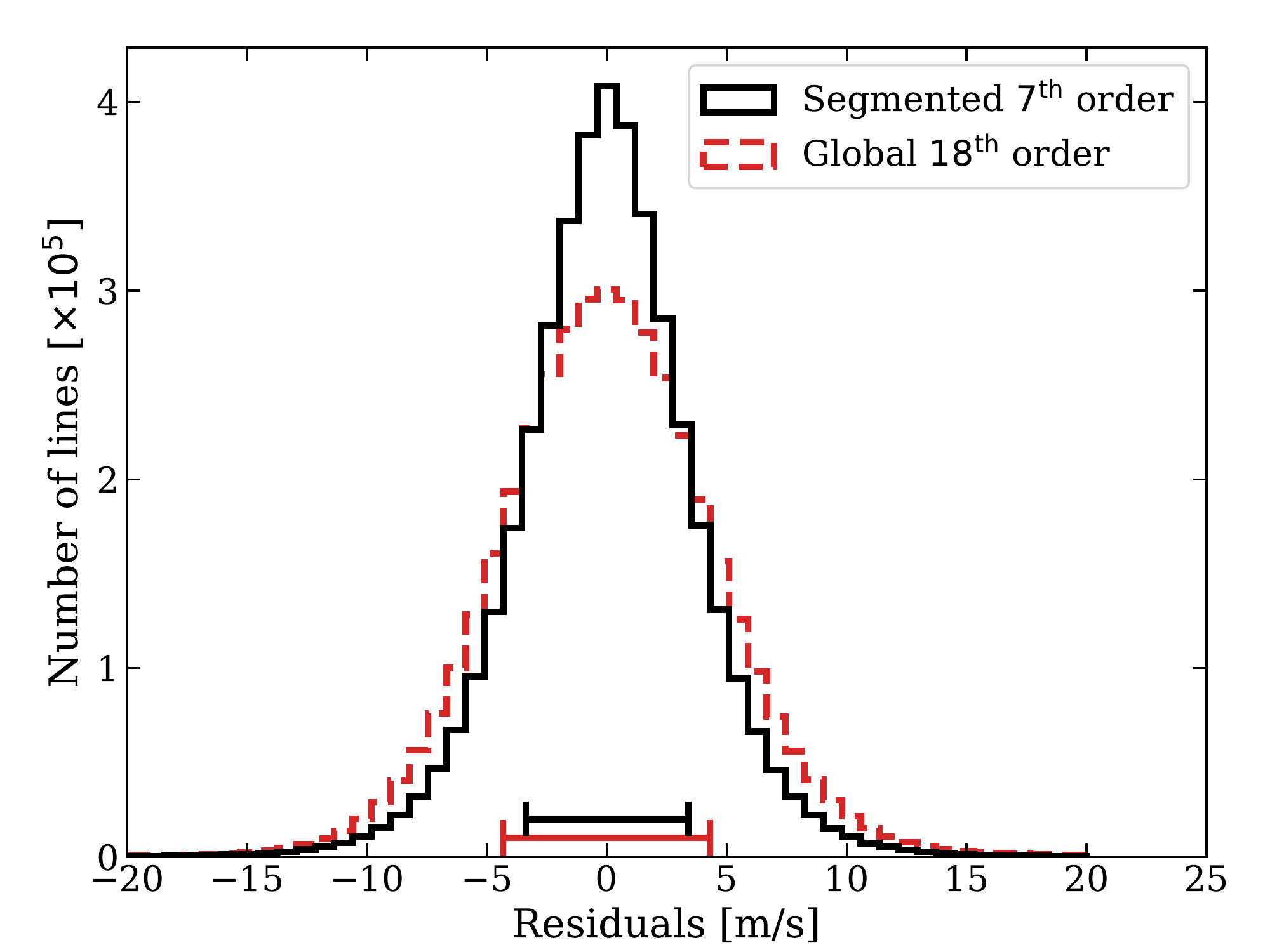}
    \caption{The histogram of residuals for the segmented $7^{\rm th}$ order polynomial in fibre A (solid black) is more centrally concentrated than that of the global $18^{\rm th}$ order polynomial (dashed red). Horizontal bars show the central 68\% of the distribution (\mps{3.4} and \mps{4.3} for the black and red, respectively). The total number of astrocomb lines is $N=4120422$.}
    \label{fig:resid_hist}
\end{figure}
\begin{figure}
    \centering
    \includegraphics[width=\columnwidth]{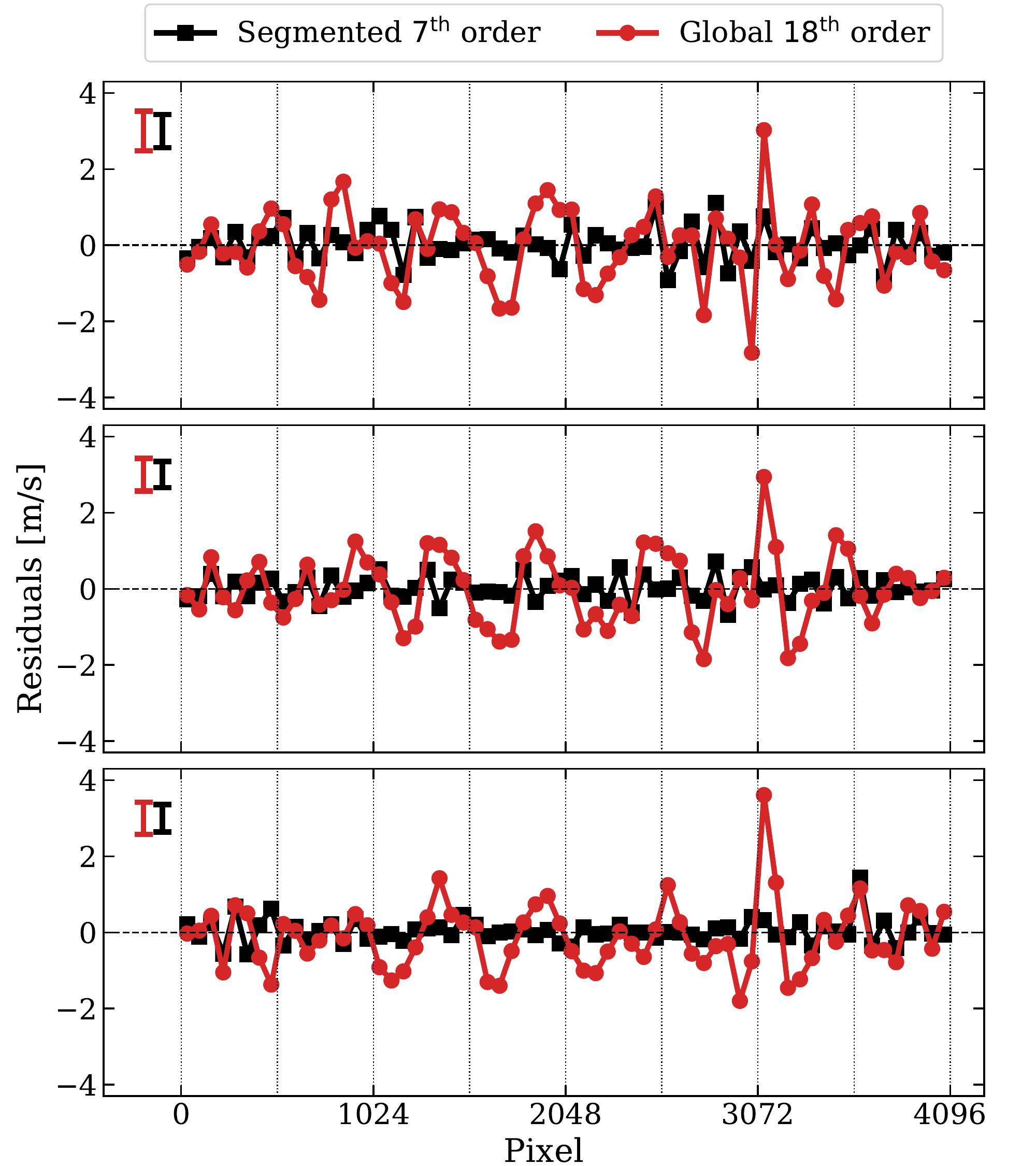}
    \caption{Mean values of the wavelength calibration residuals in 64 bins along the dispersion direction in fibre A. Points closer than 32 pixels to the segment boundaries (vertical dashed lines) have been excluded. The global polynomial fit (red line and circles) produces highly correlated residuals. 
    The structure is present and similar across all three HARPS detector blocks. The segmented order polynomial (black line and squares) shows no such structure. The vertical bars in the top left of each panel illustrate 10 times the average error on each point.}
    \label{fig:resid_structure}
\end{figure}

\begin{figure*}
    \centering
    \includegraphics[width=\textwidth]{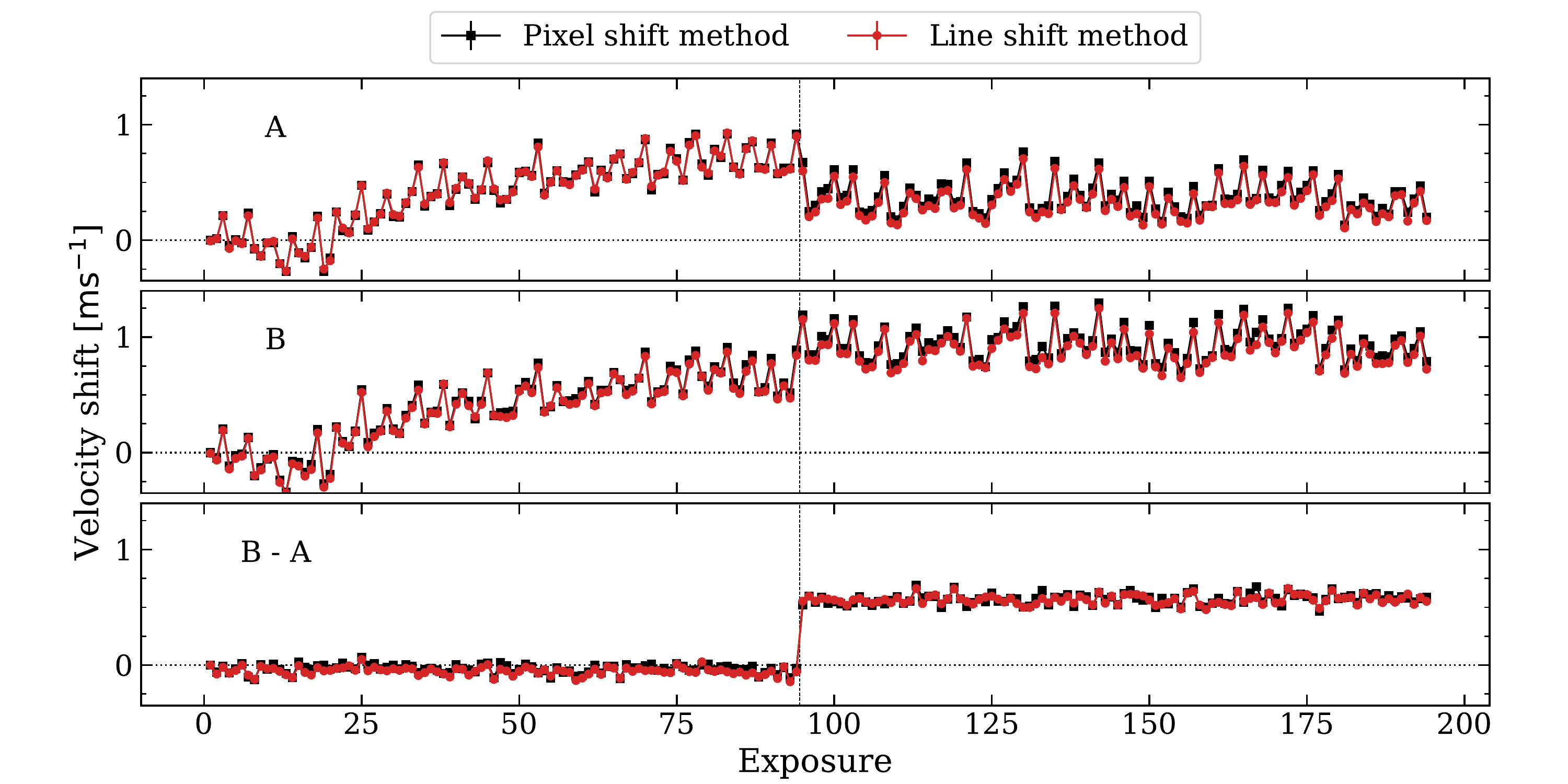}
    \caption{Velocity shift measurements for all exposures in fibres A (top panel) and B (middle panel) using the pixel shift method (black line and circles) and the line shift method (red line and squares). The velocities derived by both methods trace each other very well and are virtually indistinguishable by eye. Subtracting the spectrograph drift (B-A, bottom panel) reveals remarkable precision of each astrocomb (${\rm rms}\leqslant \cmps{4}$) but also a discrepancy in the absolute scales of the two astrocombs of $61\pm\cmps{0.6}$ level. The average photon-limited velocity precision is 
    \cmps{2.3} (\cmps{2.2}) for fibre A, \cmps{2.8} (\cmps{2.7}) for fibre B, and \cmps{3.6} (\cmps{3.5}) for B-A, using using the pixel (line) shift method.}
    \label{fig:gauss_601_uncorrected}
\end{figure*}

The histograms of the residuals for the two models are shown in \figref{fig:resid_hist}. Residuals to the segmented $7^{\rm th}$ order polynomial model are more centrally concentrated than those of the global $18^{\rm th}$ order polynomial model. The central 68\% (95\%) of residuals for the segmented $7^{\rm th}$ order polynomial model are smaller than $\mps{3.4}$ ($\mps{7.5}$). The central 68\% (95\%) of residuals for the global $18^{\rm th}$ order polynomial model are smaller than \mps{4.3} (\mps{9.1}). Therefore, the lowest AICc segmented polynomial model provides, on average, smaller residuals compared to the lowest AICc global polynomial model.

We discover that the residuals obtained from the global polynomial model exhibit a structure that correlates with pixel number. In order to more clearly illustrate the effect, we bin the residuals into 64 bins along the pixel axis (8 bins per 512-pixel wide segment) and calculate the mean and its uncertainty in each bin. The correlation pattern is similar in all 3 blocks in the y-direction on the CCD for fibre A (red circles in \figref{fig:resid_structure}). The same patterns are seen in the fibre B data (not shown). The pattern has amplitudes as high as \mps{4} with a root-mean square (rms) of $\approx \mps{1}$. To test the sensitivity of the pattern to the lines falling close to segment edges, we remove lines falling within 32 and 64 pixels around the segment boundaries and recalculate the means. We find that the pattern persists and neither its amplitude or rms changes significantly. The same result is obtained using bins of different sizes. No such correlation is seen for the segmented polynomial (black squares in \figref{fig:resid_structure}).

The pronounced residuals seen in the red circles in \figref{fig:resid_structure} are not associated with the pixel anomaly discontinuities illustrated in \figref{fig:gaps}, since these have been removed prior to polynomial fitting. However, looking at the characteristics of the continuous black line (and red squares), applying 7 offsets to move the curves together will not yield an overall trend {\it entirely} free of discontinuities. Thus we may not expect one global polynomial to provide a complete description of the data, even after correcting for the pixel anomalies. Examining figure 5 of \citet{Coffinet2019} indicates the same general phenomenon is found when using flat-fielding methods to quantify and remove the pixel anomalies. 

Whether these small remaining discontinuities are sufficient to generate the residual correlations we have found is unclear. What is clear however, is that global polynomials (i.e. a single polynomial per echelle order) should not be used to calibrate astronomical spectra used for spectroscopic velocity shift measurements. We therefore use the segmented $7^{\rm th}$ order polynomial for wavelength calibration in the rest of our analysis.

% -------------------------------------------------------------------------------------------
\section{Precision and consistency}
\label{sec:analysis}

Two types of velocity shifts are present in astrocomb time-series measurements. The first impacts on each fibre identically (e.g. velocity drifts caused by pressure and temperature variations in the spectrograph). Velocity shifts like these, common to the two fibres, can be removed using the simultaneous referencing technique \citep{Baranne1996}. The second type of velocity shift operates independently on each fibre. Quantifying the precision and consistency of the two astrocombs used in this work relies on measuring the second type of shifts in our dataset. We therefore calculate velocity shifts of all exposures in the two fibres and take their differences, effectively removing the first type of shifts. 

We measure the mean velocity of each exposure relative to the zero-point set by the first exposure in the time-series. The velocity shift is calculated using all wavelength calibrated pixels, where the shift of the $i^{\rm th}$ pixel in the $j^{\rm th}$ exposure is given by:
\begin{equation}
    \frac{\Delta v_i^j}{c} = \frac{\lambda_i^j - \lambda_i^{ref}}{\lambda_i^{ref}}.
\end{equation}
Here, $\lambda_i^{ref}$ is the $i^{\rm th}$ pixel's wavelength in the reference (first) exposure. The velocity shift of an exposure is the unweighted average of velocity shifts of all pixels. The uncertainty on the velocity shift derived above is the photon-limited velocity precision of all wavelength calibrated echelle orders, calculated using the \citet{Bouchy2001} formalism. The velocity shift of the first exposure is by definition equal to zero. We refer to this method as the ``pixel shift'' method. 

We cross-check velocity measurements of the pixel shift method using a second, independent, one. The second method, which we refer to as the ``line shift'' method, uses shifts in the astrocomb line positions on the detector to calculate the average velocity shift of an exposure. This requires a set of reference wavelength calibration coefficients: we use those of the first exposure of the series. The coefficients are used to infer wavelengths of lines in the exposure by evaluating the polynomial at the measured line positions. The velocity shift of the $i^{\rm th}$ astrocomb line in the $j^{\rm th}$ exposure is:
\begin{equation}
    \label{eq:method_line}
    \frac{\Delta v_i^j}{c} = \frac{\lambda_i^j - \lambda_i}{\lambda_i},
\end{equation}
where $\lambda_i^j$ is the inferred, and $\lambda_i$ is the true line wavelength per Equation~\eqref{eq:lfc}. The velocity shift of the exposure is the mean velocity shift of all astrocomb lines\footnote{A cut-off velocity \mps{200} was imposed to eliminate a very small number of spurious measurements (44 lines or 0.0001\% of the sample), probably associated with large line-centre uncertainties for lines with very low flux}, weighted by the errors on the inferred wavelength. The uncertainty on the velocity shift of an exposure is the standard error of the weighted mean. Because of the definition of Equation~ \eqref{eq:method_line}, the velocity shift of the first exposure is not necessarily exactly equal to zero.

The results for the entire dataset, using both methods, are plotted in \figref{fig:gauss_601_uncorrected}. The top two panels, corresponding to measurements in fibres A and B, show that spectrograph shifts are not negligible: up to about \mps{1} in the six hours of duration of the test, as measured by the shift of each fibre. However, shifts in the two fibres trace each other remarkably well, as can be seen in the bottom panel of \figref{fig:gauss_601_uncorrected}, showing their differential shift (B-A). A $\approx \cmps{60}$ velocity offset in the differential shift occurs after exposure 94, corresponding to the change from LFC2 to LFC1 in fibre B. The differential shift, B-A, should contain only those shifts that are either inherent to the two astrocomb systems or which influence the two fibres independently -- and is therefore relevant for quantifying the precision and consistency of astrocombs.

We estimate the precision of LFC1 and LFC2 as the rms of the differential shift in the LFC1 and LFC2 samples, respectively. The precision of LFC1 sample is \cmps{4.5} (\cmps{4.0})
and of LFC2 sample is \cmps{3.7} (\cmps{3.5}) using the pixel (line) shift method. This precision is $\approx 10\%$ above the average photon-limited velocity precision, which is $\cmps{3.8}$ for LFC1 and $\cmps{3.3}$ for LFC2 (see \secref{sec:data}). 

The consistency between the two astrocombs -- i.e. the jump recorded at exposure 95 in the differential shift -- is  $60.4\pm\cmps{0.6}$ for the pixel shift method, and $61.8\pm\cmps{0.6}$ for the line shift method. In order to understand the shift, one must consider that LFC1 and LFC2 have significant differences, namely different mode separations and offset frequencies. Changing from LFC2 to LFC1 is thus a major change in the calibration system akin to switching from a ThAr to a U hollow cathode lamp. Major changes in the calibration system are almost always associated with a jump in the instrumental zero-point. This implies that, in addition to the photon noise, all systematic effects associated with the wavelength calibration process will determine the consistency between LFC1 and LFC2. These include changes in the light injection into the fibres, insufficient temporal or spatial scrambling of the fibres, differences in the light path, line-spread function (LSF) variation across the detector, charge transfer inefficiency (CTI), fringing, data reduction techniques, and fitting of the data. 

\citet{ZhaoPrep} analysed data in which tests of this nature were performed in 2012 on the HARPS astrocomb prototype. Their analysis of a series of 1713 exposures shows that extreme changes to the calibration system (e.g. exchanging the photonic crystal fibre, changing the light injection, disabling the mechanical scrambler, light scrambling using the integrating sphere, mechanical realignments, etc.) produce velocity shifts with a standard deviation of \cmps{45}. Differences in illumination therefore cannot fully explain the observed \cmps{60} jump between the two astrocombs. With the exception of CTI and LSF, none of the aforementioned effects can be modelled and corrected retroactively as no suitable data were collected during the campaign.

The impact of CTI on spectroscopic velocity measurement was first measured by \citet{Bouchy2009} on the SOPHIE spectrograph. The authors of this study used a series of ThAr lamp exposures finding a clear correlation between the measured velocity shift of an exposure and its flux. Whereas shifts are as high as several tens of \mps{} at low flux ($\lesssim 600\,e^-$) observations on SOPHIE, they estimate that the effect is 2-3 times less severe on HARPS because of improved CCD performances and smaller pixels. Optimal mitigation of CTI, however, requires the acquisition of proper calibration spectra and correction of the raw frames before software post-processing, 
and is hence beyond the scope of this paper. In what follows, we use archival astrocomb observations to produce a simple model to correct flux dependent velocity shifts in HARPS spectra and apply it to our data.
As far as LSF reconstruction is concerned, we will report on our work on reconstructing the LSF of HARPS in a separate paper, with a focus on wavelength calibration accuracy \citep{MilakovicPrep}. 

\begin{figure}
    \centering
    \includegraphics[width=\linewidth]{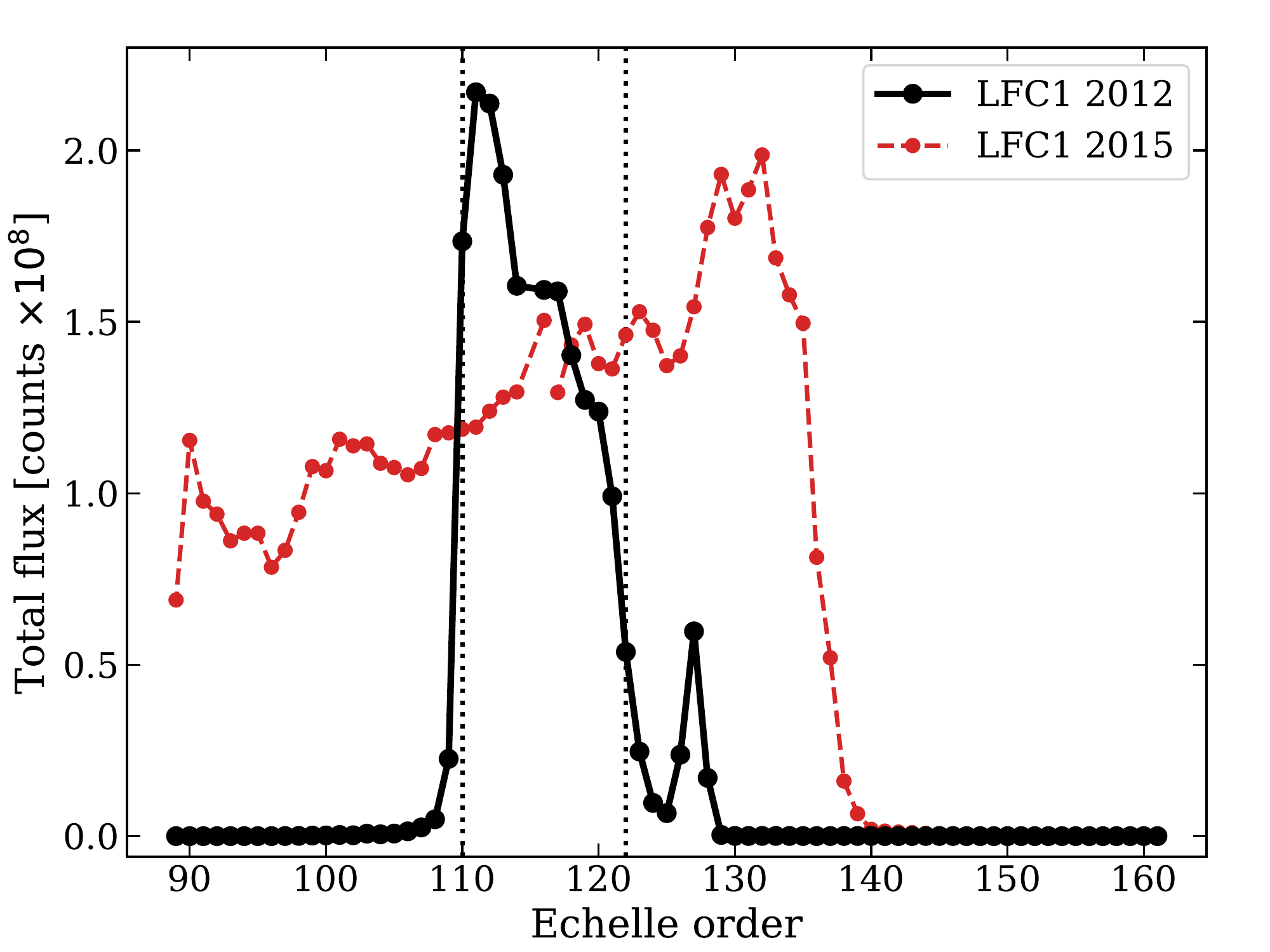}
    \caption{The wavelength range of the LFC1 prototype (2012, solid black line) is significantly shorter than that of the final design (2015, dashed red line). Our model of signal dependent velocity shifts therefore uses the average flux per line in place of the total exposure flux as in \citet{Bouchy2009}. We use echelle orders 110 to 122 (marked by the dotted vertical lines), where the flux levels are sufficiently high to be sure not to miss or falsely detect lines.}
    \label{fig:cts_per_ord_2012}
\end{figure}

\subsection{Contribution of flux dependent shifts}
\label{subsec:fluxcorr}
Following the methodology of \citet{Bouchy2009}, we look for correlations between flux in an exposure and its velocity shift using data collected during the LFC campaign of February 2012, when the LFC1 prototype was installed on HARPS.
The prototype had minor differences with respect to the final astrocomb (installed during the April 2015 campaign), the most notable being the shorter wavelength range coverage. The prototype illuminated only $\approx 33\%$ of the total wavelength range of HARPS, between $\SI{475}{\nano\meter}$ and $\SI{580}{\nano\meter}$ (echelle orders 106-128). However, we use only orders 110 to 122, in which the flux is sufficiently high to be sure of not missing or falsely detecting lines.
A comparison with the wavelength coverage of the final design is shown in \figref{fig:cts_per_ord_2012}. 

The spectral flattening unit of the astrocomb was not optimised at that time, resulting in strong and fast fluctuations in line amplitudes within a single echelle order \citep{Probst2015PhD}.
We also see a much smaller background component in the 2012 spectra, with an average B2E ratio of less than 1\%.
As noted before, the background is likely caused by the amplification of laser light before entering the photonic crystal fibre, in which the background is further amplified by non-linear processes. The background levels are lower in the 2012 data because the power in the amplification stage was significantly lower in the 2012 setup.

\subsubsection{Data and methods}
We use fifteen sets of 10 exposures each of the LFC1 prototype, taken on 15th February 2012. The first and the last sets were taken under nominal conditions, whereas the thirteen sets in between had neutral density filters of different values inserted into the light path \citep[see section 6.3 in ][]{Probst2015PhD}. The exposures were taken over a time span of 7 hours. Exposure time was $\SI{40}{\second}$ with $\SI{22}{\second}$ read-out. 

The February 2012 exposures were reduced by version v3.5 of the standard HARPS pipeline and made public via the ESO archive. We re-reduce a part of this data using pipeline version v3.8 (used to reduce the rest of our data) and find no significant differences between the products of the two pipeline versions. We therefore use the archived data (i.e. version v3.5) in this subsection. 

\begin{figure}
    \centering
    \includegraphics[width=\linewidth]{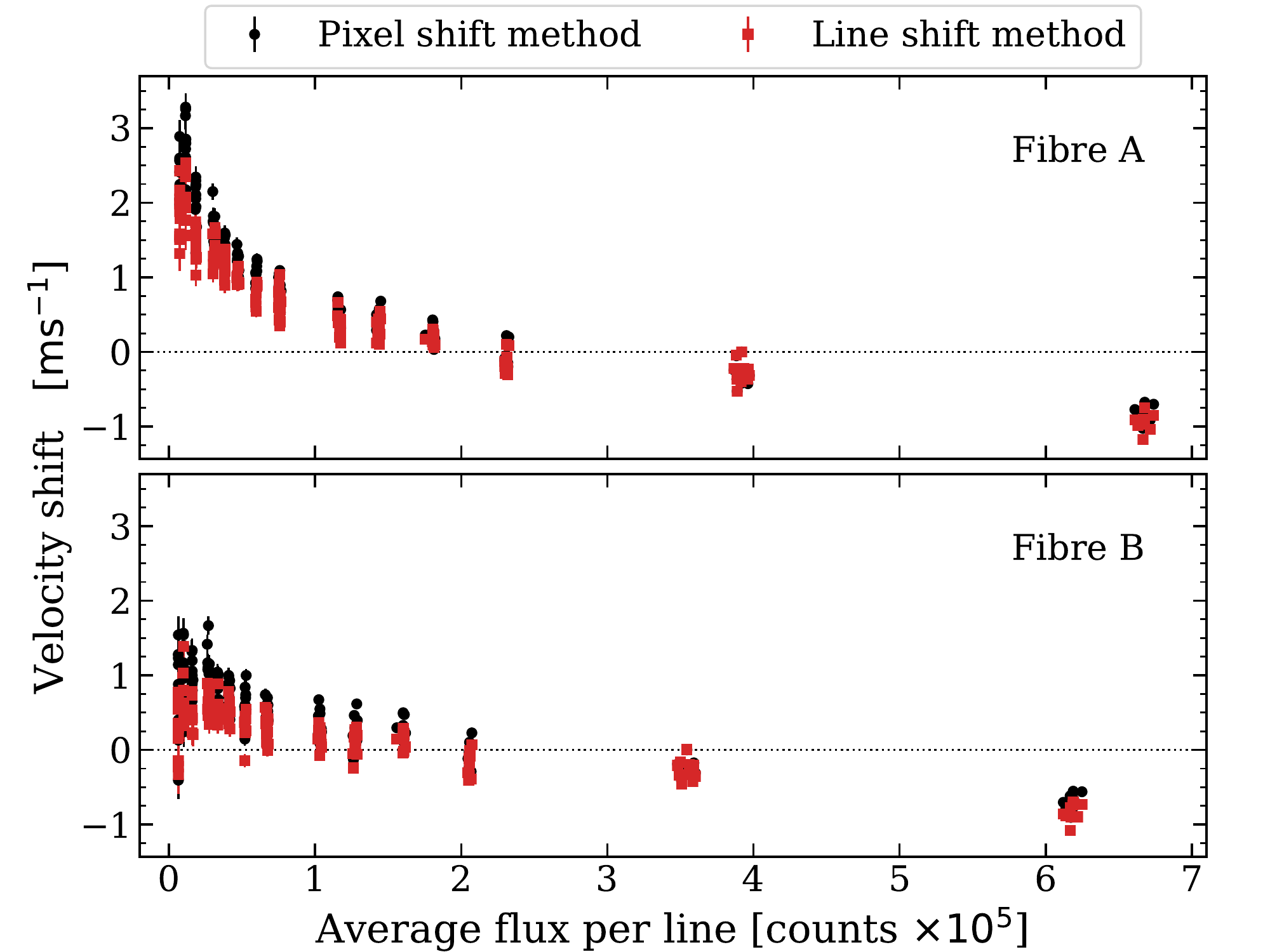}
    \caption{There is a clear trend in velocity shift with the average flux per astrocomb line. The amplitude of the shift is different for fibres A and B, as well as for the two methods we use to calculate shifts. Negative velocity shifts at the highest fluxes are due to spectrograph drifts over the duration of the series. This is because flux dependent velocity shifts are negligible for those points. This temporal shift is removed before modelling the flux dependency.}
    \label{fig:2012_flux_shift}
\end{figure}

We detect, fit, and identify all lines in echelle orders 110 to 122 in all exposures (see \secref{sec:lines}) and wavelength calibrate them (see \secref{sec:wavecal}). We then calculate velocity shifts using both the pixel shift and line shift methods described previously in this section. Given the different wavelength ranges covered by the prototype and the final design of LFC1, the total flux of an exposure is not a relevant quantity with which we can quantify flux dependent velocity shifts in the 2015 data. We therefore use the average flux per line taken over echelle orders 110 to 122. Individual line fluxes vary significantly not only across orders but also within each individual order in this data. Nonetheless, a clear trend between the average flux per line and average velocity shift is present for both fibres (\figref{fig:2012_flux_shift}).

Flux dependent velocity shifts are different for the two fibres. This is unexpected and currently not understood. This indicates that other effects, in addition to CTI, affect the velocity-shift dependency on flux. The shift to negative velocities in the last set of calibrations (without filter) are not due to flux, but spectrograph drifts with time over the duration of the test (see \figref{fig:2012_flux_shift}). Assuming a linear drift with time, we correct each exposure for the temporal component of the velocity drift by fitting a straight line to the mean observing time of the first and the last set of exposures (both without filter). We subtract this temporal drift prior to focusing on the flux dependence.

\subsubsection{Model}
\begin{figure}
    \centering
    \includegraphics[width=\linewidth]{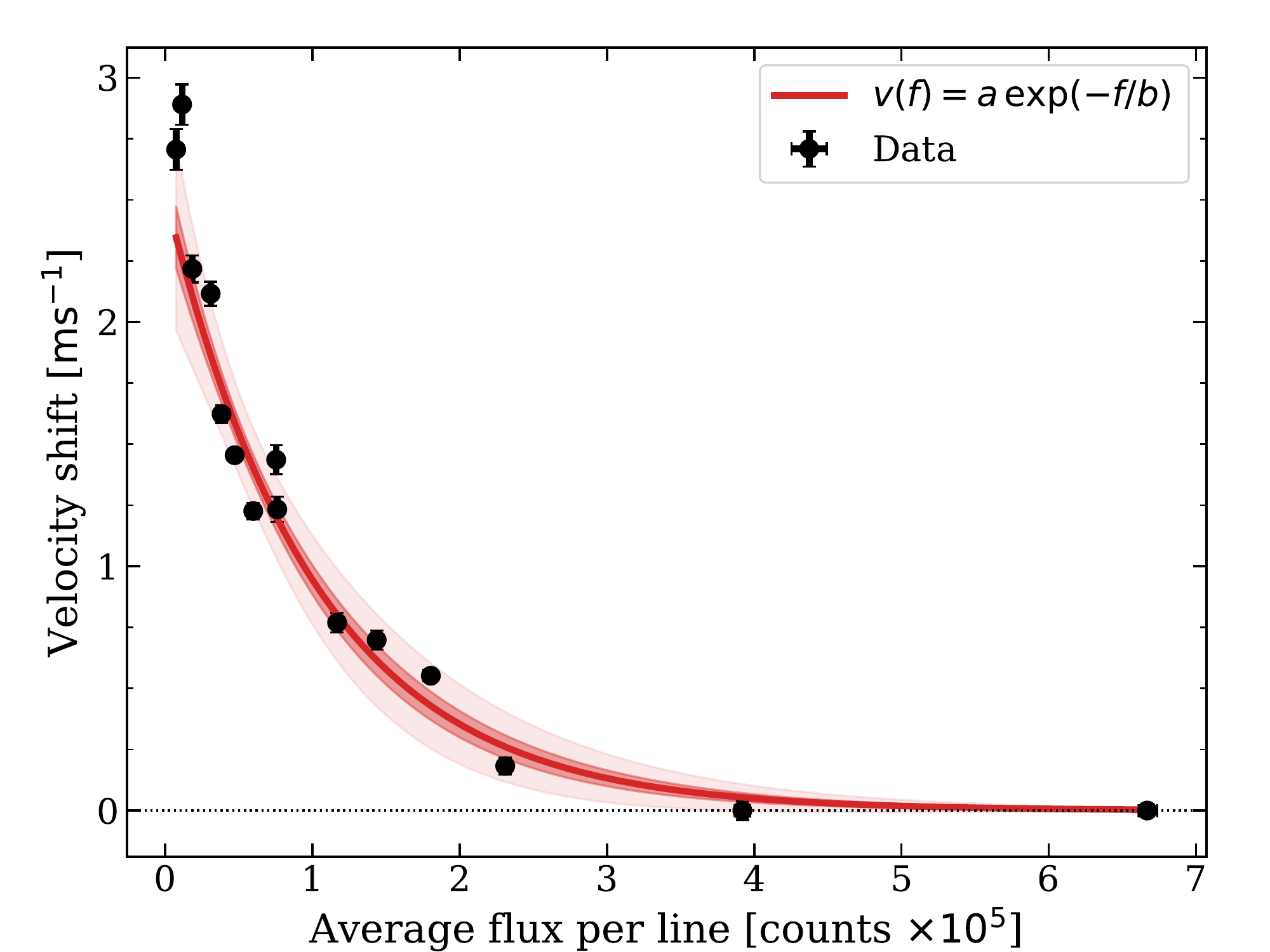}
    \caption{We fit an exponential function to the mean velocity shift of exposures taken through each neutral density filter used (black circles, 10 exposures per point) from the 2012 data, after correcting for a temporal shift component. We subsequently apply the model derived here to the 2015 data. The plot shows an example for a model in fibre A and the line shift method. The dark and light shaded areas correspond to $1\sigma$ and $3\sigma$ uncertainties on the model.}
    \label{fig:2012_model}
\end{figure}

We model the flux dependency of velocity shifts with a simple exponential model of flux:
\begin{equation}
    v(f) = a\, \exp{(-f/b)} \; [\mps{}],
\label{eq:flux_exp}
\end{equation}
where $v$ represents the velocity shift of an exposure with an average flux per line $f$. We correct each exposure for temporal drift and subsequently bin them into fifteen sets of ten prior to fitting. We use least-squares fitting to determine the values of parameters $a$ and $b$ from the data, producing four separate models: one for each combination of fibre and velocity shift measurement method (\tabref{tab:flux_model}). An example of the fit for fibre A and the line shift method is shown in \figref{fig:2012_model}. 

\begin{table}
    \centering
    \caption{Parameters of the exponential model (Equation~\ref{eq:flux_exp}) for each fibre and each velocity shift method.}
    \begin{tabular}{cccc}
    \hline
    Fibre       & {Method} &  $a\, [\mps{}]$  & $b \, [\times 10^4]$ \\
    \hline
    \multirow{2}{*}{A}  & Pixel shift  & $3.00\pm0.17$   & $9.69\pm0.90$   \\
                        & Line shift   & $2.52\pm0.15$   & $10.17\pm1.02$ \\
                                 \hline
    \multirow{2}{*}{B}  & Pixel shift  & $1.61\pm0.09$   & $11.80\pm1.25$  \\
                        & Line shift   & $1.40\pm0.10$   & $11.75\pm1.55$\\
    \hline
    \end{tabular}
   
    \label{tab:flux_model}
\end{table}

\begin{figure}
    \centering
    \includegraphics[width=\linewidth]{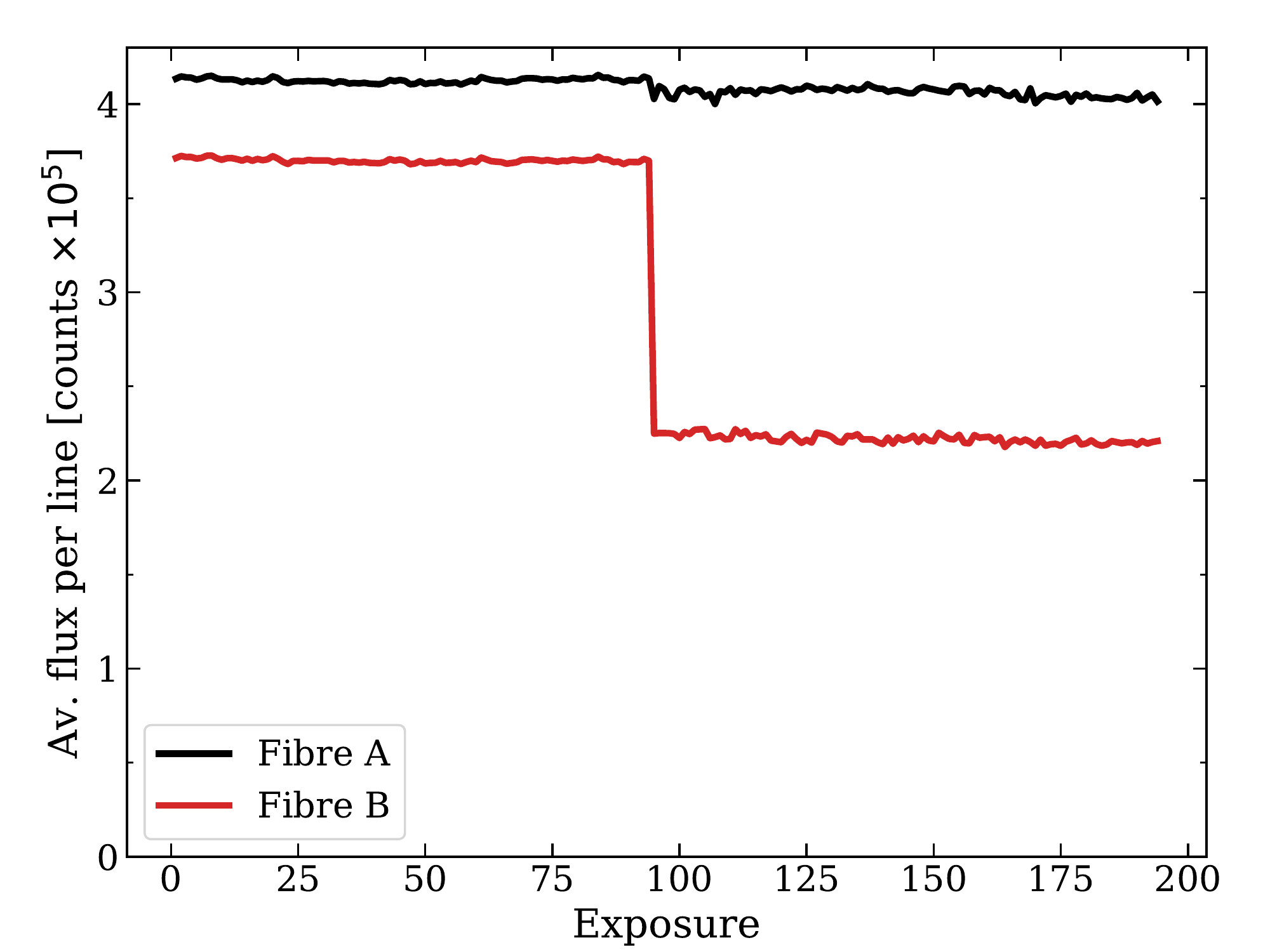}
    \caption{The average flux per line in orders 110--122 from our 2015 dataset. Fibre A (black) carried LFC2 light throughout the series. Fibre B (red) carried LFC2 light for the first 94 exposures, after which it carried LFC1 light. The average flux per astrocomb line in fibre B is 90\% of the flux in fibre A for the first 94 exposures, after which it drops to 55\%.}
    \label{fig:2015_flux}
\end{figure}

\subsubsection{Application to 2015 data}
\begin{figure*}
    \centering
    \includegraphics[width=\textwidth]{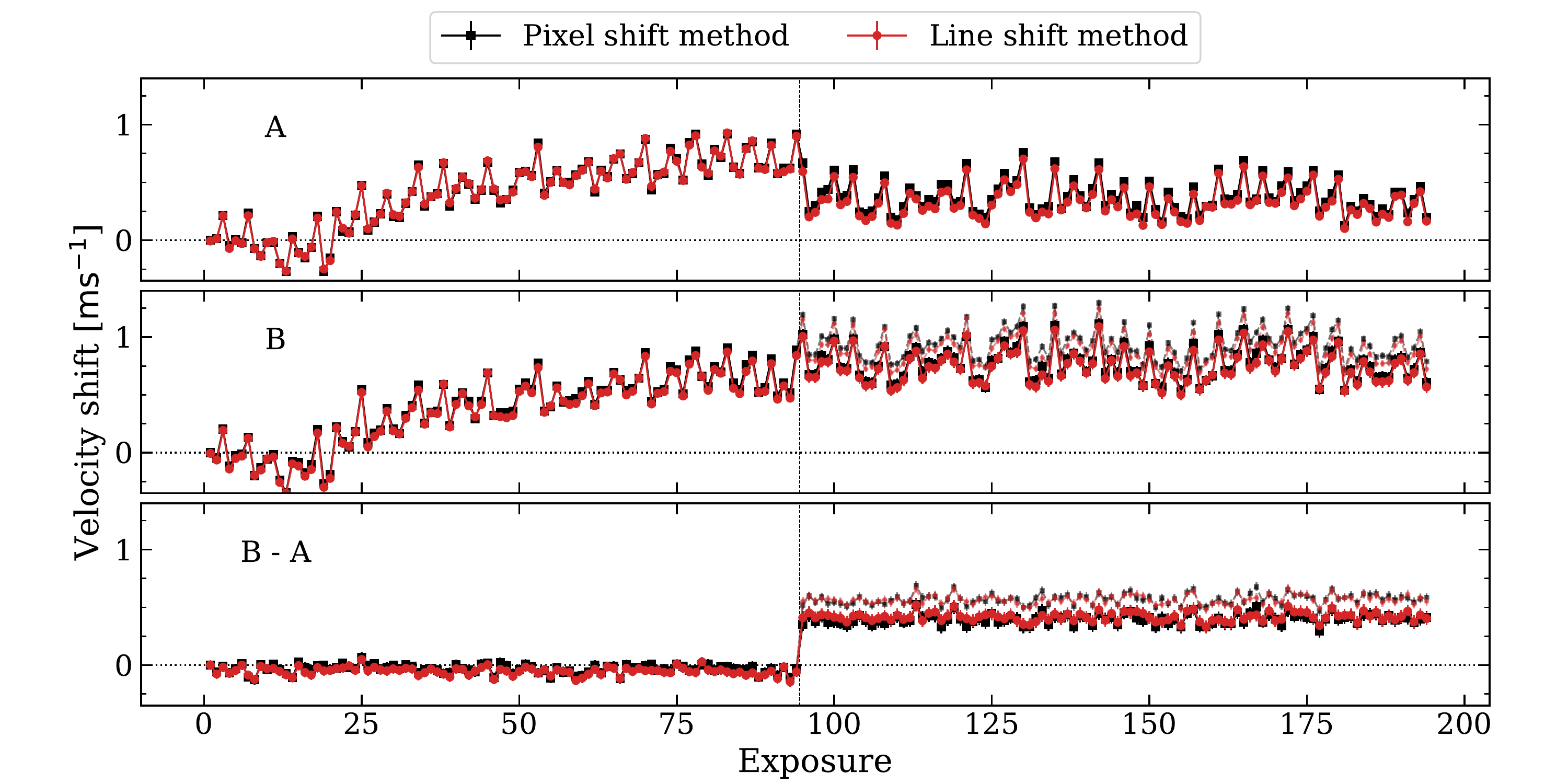}
    \caption{The same as \figref{fig:gauss_601_uncorrected}, but after correcting for flux dependent velocity shifts. This improves the consistency between LFC1 and LFC2 by $\approx 25\%$, bringing it to $\approx45\pm\cmps{0.6}$. The original, uncorrected, measurement is shown as a thin dotted line.}
    \label{fig:gauss_601_corrected}
\end{figure*}
\begin{table*}
    \centering
    \caption{A tabular overview of our astrocomb precision and consistency results. We calculate velocity shifts of astrocomb exposures for LFC1 and LFC2 samples using two independent methods. We tabulate the average photon-limited velocity precision of the sample ($\mu_{\rm PL}$, \citet{Bouchy2001}, see \secref{sec:analysis}) in column 5. The empirical precision achievable from a single exposure (rms) of the sample is tabulated in column 6. The difference between the mean velocities of each sample ($\mu$ in column 7 and the corresponding error $\sigma_\mu$ in column 8) reveal an offset in the velocity zero-point when switching from LFC2 to LFC1 in fibre B after exposure 94. The consistency between the absolute calibrations of the two astrocombs (column 9) is thus $\approx 61\pm\cmps{0.6}$. Allowing for flux-dependent velocity shifts in the data improves the consistency by $\approx 25\%$, bringing it to $\approx 45\pm\cmps{0.6}$.}

    \begin{tabular}{ccccccccc}
        \hline
        \hline
         \multirow{2}{*}{Method} & Flux & \multirow{2}{*}{Sample}  & \multirow{2}{*}{Exposures} & $\mu_{\rm PL}$ & Precision & $\mu$ & $\sigma_\mu$ &  Consistency  \\
         & corrected & & &  [\cmps{}] & [\cmps{}] &  [\cmps{}] & [\cmps{}] & [\cmps{}]  \\ 
         \hline 
         \multirow{2}{*}{Pixel shift} & \multirow{2}{*}{No}   & LFC2  & 1 - 94  & $3.4$ & $3.7$ & $ -3.2$  & $0.4$  & \multirow{2}{*}{$+60.4\pm0.6$}   \\ \vspace{0.2cm}
                                      &                       & LFC1  & 95-194  & $3.8$ & $4.5$ & $+57.2$  & $0.5$  &  \\ 
         \multirow{2}{*}{Line shift}  & \multirow{2}{*}{No}   & LFC2  & 1 - 94  & $3.3$ & $3.5$ & $ -5.2$  & $0.4$  & \multirow{2}{*}{$+61.8\pm0.6$}   \\ \vspace{0.2cm} 
                                      &                       & LFC1  & 95-194  & $3.7$ & $4.0$ & $+56.6$  & $0.4$  &  \\ 
         \multirow{2}{*}{Pixel shift} & \multirow{2}{*}{Yes}  & LFC2  & 1 - 94  & $3.4$ & $3.7$ & $ -3.2$  & $0.4$  & \multirow{2}{*}{$+43.1\pm0.6$}   \\ \vspace{0.2cm} 
                                      &                       & LFC1  & 95-194  & $3.8$ & $4.4$ & $+39.9$  & $0.4$  &  \\ 
         \multirow{2}{*}{Line shift}  & \multirow{2}{*}{Yes}  & LFC2  & 1 - 94  & $3.3$ & $3.5$ & $ -5.2$  & $0.4$  & \multirow{2}{*}{$+46.9\pm0.6$}   \\ \vspace{0.2cm} 
                                      &                       & LFC1  & 95-194  & $3.7$ & $3.8$ & $+41.7$  & $0.4$  &  \\ 
         \hline
         \hline

    \end{tabular}
    \label{tab:results}
\end{table*}
We apply the flux dependency models (\tabref{tab:flux_model}) derived from the 2012 data to the measured velocity shifts in our 2015 data. The average flux per line in the 2015 data is calculated over the same orders as are used to derive the model, plotted on \figref{fig:2015_flux} as a function of exposure number. \figref{fig:gauss_601_corrected} shows the impact of the flux correction on the measured velocity shifts. Whereas LFC2 sample velocities are mostly unchanged due to their high average flux, LFC1 sample velocities shift by $\approx\cmps{-15}$, with an uncertainty from the model of $\approx \cmps{2}$. Applying the correction improves the precision of LFC1 by $\lesssim 5\%$ and improves the consistency between LFC1 and LFC2 to $43.1\pm\cmps{0.6}$ ($46.9\pm\cmps{0.6}$) using the pixel (line) shift method. This is an improvement in absolute value of $\approx 25\%$ and is perhaps surprising given the simplicity of the model. This result demonstrates that the effect of flux on the precision and consistency of astrocomb calibrations is not negligible. This strongly motivates the importance of detailed measurements and parameterisation of such dependencies when attempting astronomical calibrations and observations approaching $\approx\cmps{1}$.

\subsection{Achievable precision}
\label{subsec:precision}

\begin{figure}
    \centering
    \includegraphics[width=\linewidth]{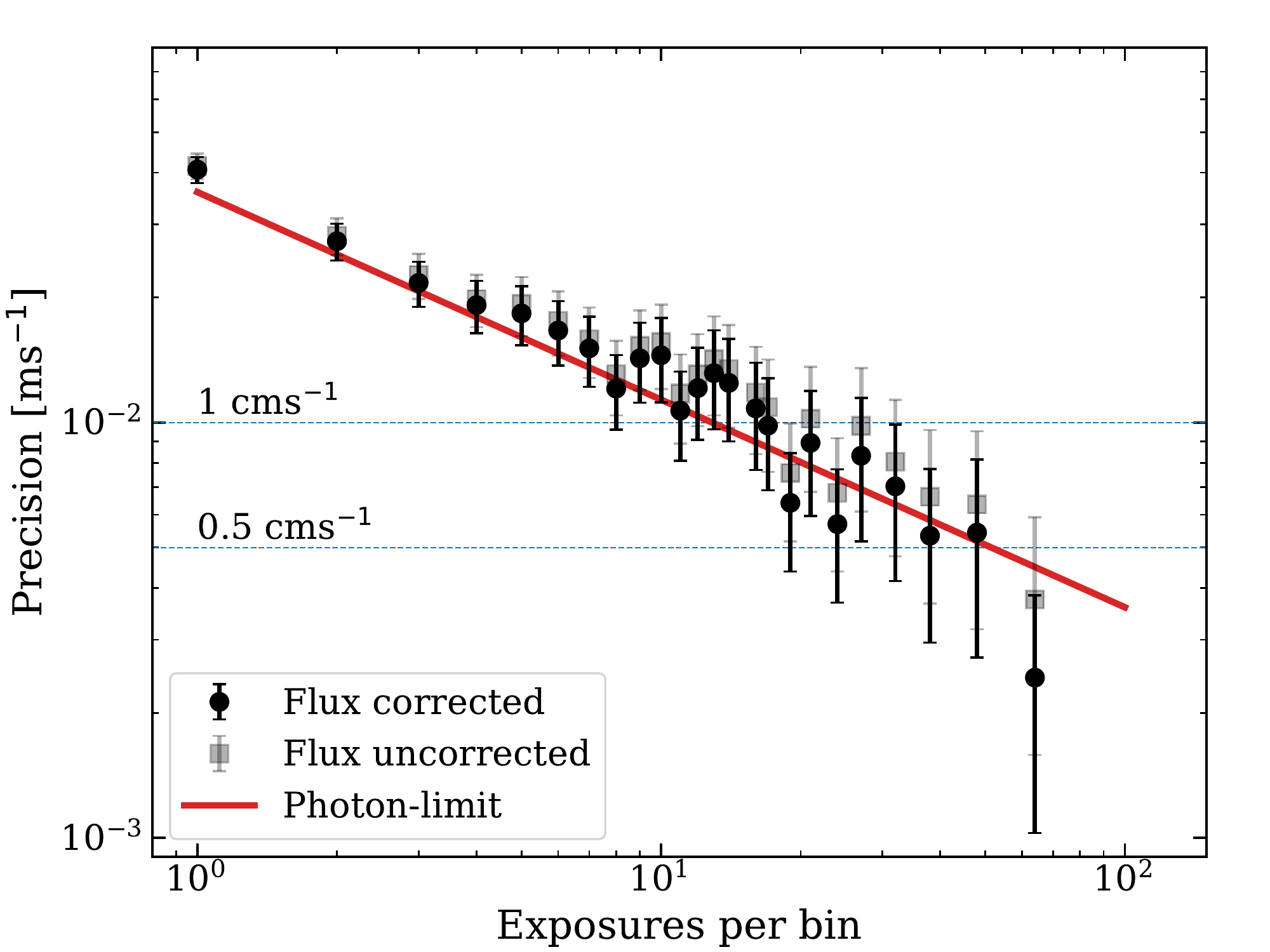}
    \caption{The calibration precision of astrocomb velocity measurements (as measured by the standard deviation) improves as we bin the measurements in bins of increasing size in a way that is in agreement with the photon-limited precision (red line). The precision is additionally improved after correcting for flux dependent effects. }
    \label{fig:precision}
\end{figure}

In applications such as extrasolar planet radial velocity measurements, it is of interest to explore the limiting calibration precision achievable in a reasonable observing time. To examine this, we begin by taking the data illustrated in \figref{fig:gauss_601_corrected} but remove the break in mean velocity shift seen at exposure 94. This is done simply by calculating the means either side of exposure 94 (see \tabref{tab:results}) and removing both i.e. normalising to zero means either side of exposure 94. The data have previously been flux-corrected as described in \secref{subsec:fluxcorr} to account for the different flux levels of the two astrocombs. The rms velocity shift is then calculated for all 194 points.  We then bin the number of exposures in increasingly large bins, starting with 2 exposures per bin, and increasing the number of points per bin. This is illustrated in \figref{fig:precision}, which shows that for maximal binning, a radial velocity calibration precision $\approx 0.5$ cm\,s$^{-1}$ can in principle be achieved. Comparing this with the expectation based on the photon-limited velocity precision (continuous red line in \figref{fig:precision}), we see good agreement between the theoretical prediction and observations. 

The 194 exposures used for the procedure above correspond to a total integration time of 1.6 hours (taken over a period of 6 hours - see \secref{sec:data}). We conclude from this that, given the corrections applied above, a realistic achievable calibration precision is of order \cmps{1}.

% -------------------------------------------------------------------------------------------
\section{Results}
\label{sec:results}

Our main results are:
\begin{enumerate}

\item Using global polynomials for wavelength calibration produces residuals which correlate with pixel number (see \figref{fig:resid_structure}) even when high ($18^{\rm th}$) order polynomials are used. The pattern is highly modulated, has amplitudes as high as \mps{4}, and is present across the detector. 

\item We find absolute velocity shifts between the wavelength solutions measured using two independent astrocombs of $\approx 61\pm\cmps{0.6}$ when only HARPS instrumental drifts are removed. This unexpected result can be partly attributed to differences between the two astrocomb flux levels, since a clear (non-linear) correlation is seen between the mean flux and a mean shift velocity in spectral line positions (\figref{fig:2012_model}). However, even allowing for this, a significant absolute shift remains: $\approx 45\pm\cmps{0.6}$ or $\Delta\lambda/\lambda = 1.3\times 10^{-9} $ (\figref{fig:gauss_601_corrected}).

\item The precision of each astrocomb in a single exposure is $\lesssim \cmps{4}$ ($\approx 10\%$ higher than the measured photon-limited precision). Precision remains unchanged when a different comb is injected in the second fibre demonstrating it remains unchanged by using two independent systems. A realistic achievable velocity calibration precision is of order \cmps{1} provided systematics are carefully measured and removed (see \secref{subsec:precision}).

\end{enumerate}

% -------------------------------------------------------------------------------------------

\section{Discussion}
\label{sec:discussion}

We set out to determine the limiting precision with which spectroscopic velocity shifts in high resolution spectra can be measured using current methodology, in the context of the science goals motivating the construction of future large optical observatories. In this sense, we have achieved our goals. 
Firstly, we demonstrated that astrocombs can achieve repeatabilities of around \cmps{1} using advanced methods in conjunction with a second, independent, astrocomb. We thus go beyond the results of \citet{Wilken2012} where only one astrocomb was used. Secondly, by comparing measurements from two independent astrocombs, we discovered unexpected and substantial wavelength zero-point offsets between astrocombs, the causes of which are not yet completely understood.

We identify systematics introduced into the astrocomb measurements by the combined effects of the detector morphology, the CTI during data read-out, and imperfect LSF modelling (i.e. the Gaussian approximation). We have not identified any systematic effects associated with the astrocomb itself. This implies that improvements in the former three will enable precision improvements that approach theoretical limits. A tunable astrocomb, capable of scanning the full separation between two astrocomb modes would be ideal to better understand the system. Astrocombs with large ($\approx \SI{10}{\giga\hertz}$) native mode separations should be able to provide such a feature, e.g. those based on electro-optic combs \citep{Obrzud2019}. 

It was only possible to quantify the zero-point offset resulting from the change of the astrocomb because two astrocombs were used simultaneously on HARPS. This would not have been possible if only one astrocomb had been used, but instead the first astrocomb had simply been replaced by a second. This point merits careful consideration when designing astronomical measurements requiring long-term stability. Whilst observing a set of radial velocity standards before and after the astrocomb change might be sufficient for exoplanet detection studies, achieving the stability required for the redshift drift measurement warrants a different approach.

Lastly, we discovered highly correlated wavelength residuals resulting from employing global polynomials for wavelength calibration -- the default method in essentially all previous echelle spectroscopy. The discovery was made whilst investigating different calibration algorithms (global versus segmented polynomials, \secref{sec:wavecal}) and could only be made due to the large number of astrocomb lines available. 

The expected effect of the correlated residuals is to introduce spurious velocity shifts in the data. The severity of this effect depends both on the science goal of observations and on individual characteristics of the target: the number of useful lines and where they fall with respect to the correlated structure.
For example, the most precise redshift measured to date for any single heavy element absorption line at high redshift, using optical spectroscopy, has a redshift uncertainty of around $4 \times 10^{-6}$, or around \mps{5}. If global polynomials are used, correlated calibration residuals may emulate a varying fine structure constant at a level $\Delta\alpha/\alpha$ of around $10^{-6}$ in this single line. This is of the same order as the statistical error in this system.
Radial velocity measurements from stellar spectra will be influenced in a similar way, where the signal could emulate periodicity. The exact period of the spurious signal will depend on time sampling of observations in addition to which lines are used for the measurement. Therefore, the correlated residuals also have the potential to emulate spurious exoplanet detections. 
Finally, the expected signal in the redshift drift measurement is of order \cmps{1} \citep[see Fig. 2 in][]{Liske2008}. Correlated residuals at the level of $\mps{4}$ would therefore render detection of redshift drift impossible. However, the results presented here are rather encouraging: provided segmented polynomials are used (assuming existing technology), the calibration precision of $\approx \cmps{1}$ has now just about been reached.

\section*{Acknowledgements}
Based on observations collected at the European Southern Observatory under ESO programme ID 60.A-9036. We thank the members of the HARPS astrocomb commissioning team who operated the astrocomb, collected and processed the data. In particular, DM thanks Rafael A. Probst for providing information about the astrocombs and helpful discussion. We also thank the referee for their useful comments. The analysis within this paper made extensive use of \texttt{scipy} \citep{Jones2001}, \texttt{numpy} \citep{Walt}, and \texttt{cfitsio} \citep{Pence1999}. 

%%%%%%%%%%%%%%%%%%%%%%%%%%%%%%%%%%%%%%%%%%%%%%%%%%



\bibliographystyle{mnras}
\bibliography{main} % if your bibtex file is called example.bib

%\begin{thebibliography}{99}
%\end{thebibliography}

%%%%%%%%%%%%%%%%%%%%%%%%%%%%%%%%%%%%%%%%%%%%%%%%%%

%%%%%%%%%%%%%%%%% APPENDICES %%%%%%%%%%%%%%%%%%%%%
\appendix

\section{Mode identification issue with LFC1 in our dataset}
\label{app:harps_shift}
During our analysis we noticed a systematic velocity shift of approximately \mps{45} between the LFC1 and LFC2 wavelength calibrations and between LFC1 and attached ThAr calibrations. A shift of this magnitude cannot be explained by spectrograph drifts, which made us suspect we have misidentified an astrocomb line. \citet{Coffinet2019} used the same dataset in their analysis and noted that the offset frequency of LFC1 was probably different by $\SI{100}{\mega\hertz}$ from that reported in the observing log. It is likely that the change in the offset frequency was made by the operator and not noted in the system. In what follows, we provide a definitive proof of the shift's existence and measure its value. 

\begin{figure}
    \centering
    \includegraphics[width=\linewidth]{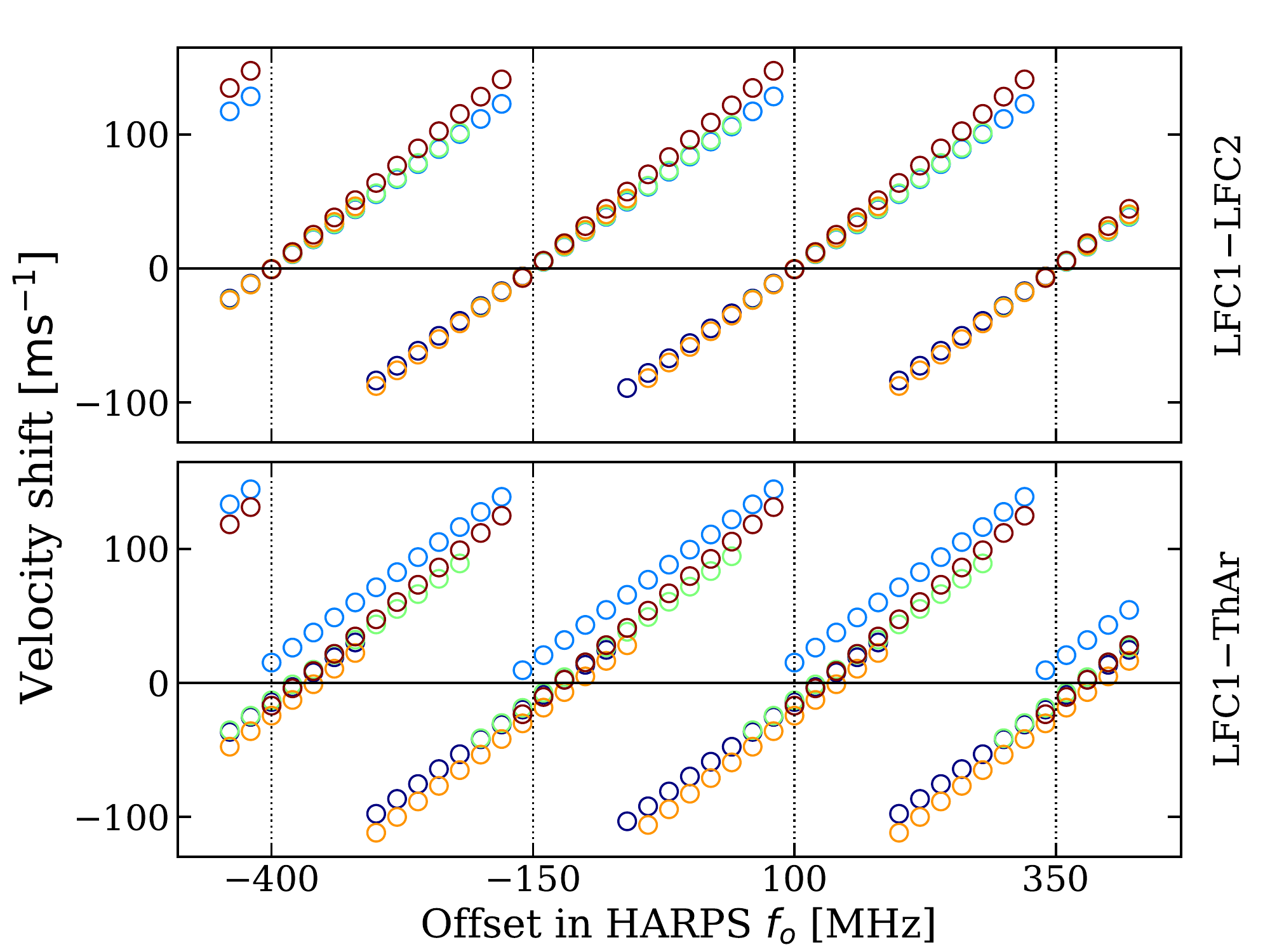}
    \caption{We manually change the offset frequency of LFC1 in steps of \SI{20}{\mega\hertz} and calculate the average velocity shift with respect to LFC2 (top panel) and ThAr (bottom panel) wavelength calibrations in several echelle orders (different colours). We find that LFC1 offset frequency needs to be shifted by $\SI{100}{\mega\hertz}$ from what was reported in the observing log in order to be consistent with the ThAr and LFC2 calibrations. This is probably due to logging error.}
    \label{fig:anchor_offset}
\end{figure}

We return to the mode identification algorithm and perform an exercise to verify that we are indeed assigning modes correctly. To this end, we use LFC1, LFC2, and ThAr spectra taken within a short time period from each other to ensure spectrograph drifts are small. We select several echelle orders in the ThAr exposure and wavelength calibrate them ourselves by fitting a third order polynomial through several ThAr lines for which wavelengths are tabulated in the \citet{Palmer} atlas. We then identify, by eye, LFC1 and LFC2 lines that coincide with ThAr lines to within 1 pixel on the detector. We now take those particular astrocomb lines (one LFC1 and one LFC2 per order) and estimate their mode number from the wavelength of the coinciding ThAr line. Knowing the modes of the astrocomb lines, we use them to wavelength calibrate each echelle order as described in \secref{sec:wavecal}, but we change the frequency of LFC1 lines during fitting in steps of $\SI{20}{\mega\hertz}$ in the range $\SI{\pm 440}{\mega\hertz}$. For each frequency step, we calculate the average velocity shift between ThAr, LFC1, and LFC2 wavelength calibrations using the pixel shift method (see \secref{sec:analysis} for details). We find that the LFC1 offset frequency needs to be shifted by $\Delta f_o = \SI{100}{\mega\hertz}\pm n\times \SI{250}{\mega\hertz}$ (with $n$ an integer number) in all echelle orders examined to achieve agreement between all three calibrations (see \figref{fig:anchor_offset}). The frequency shifts are degenerate with $\SI{250}{\mega\hertz}$, which is the repetition frequency of the LFC1 spectrum before mode filtering \citep[see ][for more details]{Probst2020}. 

Assuming the smallest possible shift, we change the LFC1 offset frequency by $\Delta f_o =\SI{+100}{\mega\hertz}$ during mode identification step of our analysis, i.e. Equation~ \eqref{eq:lfc}. The LFC1 offset frequency in  \tabref{tab:setup} (\SI{5.7}{\giga\hertz}) already reflects this change.

%%%%%%%%%%%%%%%%%%%%%%%%%%%%%%%%%%%%%%%%%%%%%%%%%%

% Don't change these lines
\bsp	% typesetting comment
\label{lastpage}
\end{document}